\documentclass[11pt,twoside,a4paper]{article}

\usepackage[pdftex]{graphicx} 
\usepackage{amsmath,amssymb,amsthm}
\usepackage[portuges]{babel}
\usepackage[ansinew]{inputenc}
\usepackage{a4wide}
\usepackage{xspace}

\newcommand{\dd}{\mathrm{d}}

\newtheorem{theorem}{Teorema}

\begin{document}

\title{Uma forma bidimensional que maximiza\\a resistência aerodinâmica newtoniana}

\author{Paulo D. F. Gouveia\small ${\,}^*$ \normalsize\\
        \texttt{pgouveia@ipb.pt} \and
        Alexander Plakhov\small ${\,}^{**}$ \normalsize\\
        \texttt{plakhov@mat.ua.pt} \and
        Delfim F. M. Torres\small ${\,}^{**}$ \normalsize\\
        \texttt{delfim@ua.pt}}

\date{\small ${}^*$Escola Sup. Tecn. e Gestão, Instituto Politécnico de Bragança, 5301-854 Bragança\\
${}^{**}$Departamento de Matemática, Universidade de Aveiro, 3810-193 Aveiro}

\maketitle

\vspace{-1.0cm}
\renewcommand{\abstractname}{}
\begin{abstract}
\textbf{Abstract}.
In a previous work \cite{Plakhov07:CM, Plakhov07} it is investigated, by means of computational
simulations, shapes of nonconvex bodies that maximize resistance to its
motion on a rarefied medium, considering that bodies are moving forward
and at the same time slowly rotating. Here the previous results are improved:
we obtain a two-dimensional geometric shape that confers to the body
a resistance very close to the supremum value ($R=1.4965<1.5$).
\end{abstract}

\renewcommand{\abstractname}{Resumo}
\begin{abstract}
Um corpo bidimensional, apresentando um ligeiro
movimento rotacional, desloca-se num meio rarefeito de partículas
que colidem com ele de uma forma perfeitamente elástica.
Em investigações que os dois primeiros autores realizaram anteriormente \cite{Plakhov07:CM, Plakhov07},
procuraram-se formas de corpos que maximizassem a força de travagem do meio ao seu movimento.
Dando continuidade a esse estudo, encetam-se agora novas investigações que
culminam num resultado que representa um grande avanço qualitativo relativamente aos então alcançados. Esse resultado, que agora se apresenta, consiste numa forma bidimensional que confere ao corpo uma resistência muito próxima do seu limite teórico. Mas o seu interesse  não se fica pela maximização da resistência newtoniana; atendendo às suas características, apontam-se ainda outros domínios de aplicação onde se pensa poder vir a revelar-se de grande utilidade.
Tendo a forma óptima encontrada resultado de estudos numéricos, é objecto de um estudo adicional de natureza analítica, onde se demonstram algumas propriedades importantes que explicam em grande parte o seu virtuosismo.
\end{abstract}

\medskip


\smallskip

\textbf{Palavras chave:} corpos de resistência máxima, bilhares,
problema aerodinâmico newtoniano, retrorreflector.

\medskip


\smallskip


\medskip


\section{Introdução}
Uma área de investigação da Matemática contemporânea ocupa-se com a procura de formas de corpos, dentro de classes predefinidas, que permitam minimizar ou maximizar a resistência
a que ficam sujeitos quando se desloquem em meios rarefeitos.
O primeiro problema desta natureza remonta já à década de $80$ do século XVII,
altura em que Isaac Newton estudou, em \cite{newton1686},
um problema de resistência mínima para uma classe específica de corpos convexos,
que se deslocassem em meios de partículas infinitesimais, de tal modo rarefeitos que fosse possível negligenciar qualquer interacção entre as partículas, e que a interacção destas com o corpo pudesse ser descrita por colisões perfeitamente elásticas.
Mais recentemente temos assistido a desenvolvimentos importantes
nesta área com a generalização do estudo a novas
classes de corpos e a meios com características menos restritivas:
problemas de resistência em corpos não
simétricos~\cite{buttazzo95,buttazzo97,buttazzo93,robert06,robert01b},
em corpos não convexos de colisões singulares~\cite{brock96,buttazzo93,comte01,robert01}
e de colisões múltiplas~\cite{Plakhov03b,plakhov03,Plakhov04},
corpos de superfície desenrolável~\cite{robert01},
considerando colisões com atrito~\cite{horstmann02}
e em meios de temperatura positiva~\cite{Plakhov05}.
Mas, quase invariavelmente, os resultados que têm vindo a ser publicados têm dado especial atenção a classes de corpos convexos.

A convexidade de um corpo é
uma condição suficiente para que a resistência seja função unicamente de colisões singulares
--- todas as partículas colidem uma só vez com o corpo.
Esse atributo permite reduzir consideravelmente
a complexidade dos problemas tratados. Mesmo os vários estudos sobre classes de corpos não convexos que têm surgido, especialmente na última década, assentam quase sempre
em condições que garantem um único impacto por partícula ---~\cite{brock96,buttazzo93,comte01,robert01}.
Só muito recentemente, começaram a surgir alguns estudos prevendo múltiplas
reflexões, como é o caso dos trabalhos de Plakhov~\cite{Plakhov03b,plakhov03,Plakhov04}.

Na classe de corpos convexos, o problema reduz-se normalmente 
à minimização da funcional de Newton --- uma fórmula analítica
para o valor da resistência, proposta por Isaac Newton \cite{newton1686}.
Mas no contexto de corpos não convexos não existe
qualquer fórmula simples conhecida para o cálculo da resistência.
Ainda que seja extremamente complexo, em geral, tratar analiticamente
problemas de múltiplas colisões, para alguns problemas de minimização específicos
a tarefa não se tem revelado particularmente difícil, existindo inclusive
já alguns resultados disponíveis~\cite{Plakhov03b,plakhov03}.
Se, pelo contrário, considerarmos o problema de maximização,
então a solução chega mesmo a ser trivial --- para qualquer dimensão, basta que a parte frontal
do corpo seja ortogonal à direcção do movimento.

E se o corpo exibir, para além do
seu movimento de translação, um ligeiro movimento rotacional?
Quando pensamos neste tipo de problemas temos, por exemplo, em mente
satélites artificiais, de órbitas relativamente baixas,
que não disponham de qualquer sistema de controlo que
estabilize a sua orientação, ou outros engenhos em
condições semelhantes. Nessa situação, imaginamos
que, ao longo do seu percurso, o engenho rode
lentamente sobre si próprio.

O problema de minimização
da resistência média em corpos rotativos não convexos foi já estudado
para o caso bidimensional~\cite{Plakhov04}: demonstrou-se
que a redução da resistência que é possível obter em relação
ao caso convexo não ultrapassa os $1.22\%$. Por sua vez,
o problema de maximização da resistência média de corpos em rotação
está longe de ser trivial, contrariamente ao que se
passa quando se trata de movimento puramente translacional.
Foi, por isso, esta classe de problemas o objecto de estudo do trabalho realizado pelo autores em \cite{Plakhov07:CM, Plakhov07}:
investigaram-se formas de corpos não convexos que maximizassem a resistência que os mesmos teriam que enfrentar quando se deslocassem em meios rarefeitos e, simultaneamente,
exibissem um ligeiro movimento rotacional.
Com o estudo numérico que se levou a cabo, foram encontradas
várias formas geométricas que conferiam aos corpos valores de resistência
bastante interessantes; mas foi em investigações posteriores, desenvolvidas na continuação desse trabalho, que os autores conseguiram chegar ao melhor dos resultados: uma forma bidimensional que confere ao corpo uma resistência muito próxima do seu limite máximo teórico. É este último resultado
que agora aqui se apresenta.

A apresentação do trabalho encontra-se organizada como se segue.
Em \S\ref{sec:defBidimens} começamos por definir, para o caso bidimensional, o problema de maximização, objecto do presente estudo.
Depois, em \S\ref{sec:estNumer}, descrevemos o estudo numérico que foi realizado no encalço
do corpo de resistência máxima e apresentamos o principal resultado original deste trabalho: uma forma bidimensional que maximiza a resistência newtoniana.
A forma bidimensional encontrada é então objecto de um estudo em \S\ref{sec:caract} onde se demonstram algumas propriedades importantes que ajudam a explicar o valor de resistência que apresenta.
Em \S\ref{sec:outrasAplic} apresentamos um estudo exploratório sobre outras possíveis aplicações do nosso resultado.
Por fim, em \S\ref{sec:concl}, apresentamos as principais conclusões do presente trabalho e incluímos alguns apontamentos sobre as possíveis direcções do trabalho a realizar futuramente.


\section{Definição do problema para o caso bidimensional}
\label{sec:defBidimens}
Considere-se um disco em rotação lenta e uniforme,
deslocando-se numa direcção paralela ao seu plano.
Denotemos o disco de raio $r$ por $C_r$ e a sua fronteira por $\partial C_r$.
Retiremos então pequenas porções do disco ao longo de todo o seu perímetro,
numa vizinhança $\varepsilon$ de $\partial C_r$, com $\varepsilon \in \mathbb{R}_+$
de valor arbitrariamente pequeno quando comparado com o valor de $r$.
Ficamos assim com um novo corpo $B$ definido por um subconjunto de $C_r$ e caracterizado por uma certa rugosidade ao longo de
todo o seu perímetro. 
A questão essencial que se coloca é a seguinte:
até quanto pode ser aumentada a resistência de um corpo $B$?
Mais do que conhecermos o valor absoluto dessa resistência, estamos
sobretudo interessados em saber qual o ganho que se consegue
obter em relação à resistência do corpo liso
(contorno perfeitamente circular, neste caso), ou seja,
conhecer o valor normalizado
\begin{equation}
\label{eq:normalizacao}
R(B)=\frac{\text{Resistência}(B)}{\text{Resistência}(C_r)} \text{.}
\end{equation}
É possível, desde logo, conhecermos alguns valores de referência importantes
para a resistência normalizada:
$R(C_r)=1$ e o valor da resistência $R(B)$ terá que se situar
entre $0.9878$ (\cite{Plakhov04}) e $1.5$.
O valor $1.5$ será hipoteticamente atingido se todas as partículas forem reflectidas pelo
corpo com uma velocidade 
$\mathbf{v}^+$ (velocidade com que as partículas se afastam definitivamente do corpo)
oposta à velocidade de incidência $\mathbf{v}$ (velocidade com que as partícula
atingem o corpo pela primeira vez), $\mathbf{v}^+=-\mathbf{v}$, situação em que
é transmitida ao corpo a máxima quantidade de movimento.
É ainda possível conhecermos o valor da resistência de alguns corpos
elementares do tipo $B$. É o caso, por exemplo, de discos com o contorno integralmente formado por reentrâncias rectangulares arbitrariamente pequenas ou com a forma de triângulos rectângulos isósceles.
Tal como demonstrado em \cite{Plakhov07:CM, Plakhov07}, esses corpos têm associado uma resistência, respectivamente, de $R=1.25$ e $R=\sqrt{2}$.

Para além de ser definido numa vizinhança $\varepsilon$ interior da fronteira do disco $C_r$, assume-se que o corpo que se pretende maximizar 
é um conjunto $B\in\mathbb{R}^2$ conexo e limitado, e
com fronteira $\partial B$ seccionalmente suave. Considere-se então um bilhar em $\mathbb{R}^2\setminus B$.
Uma partícula infinitesimal move-se livremente, até que, ao colidir com o corpo $B$, sofre várias reflexões (uma no mínimo) em pontos regulares da sua fronteira $\partial B$, acabando por retomar um movimento livre que a afasta definitivamente do corpo.
Represente-se por $\text{conv}B$ o conjunto definido pelo invólucro convexo de $B$.
A partícula intercepta por duas vezes o contorno $\partial(\text{conv}B)$:
quando entra no conjunto $\text{conv}B$ e no momento em que o
abandona.
Considere-se $L=|\partial(\text{conv}B)|$ o comprimento
total da curva $\partial(\text{conv}B)$ e 
represente-se por $\mathbf{v}$ e $\mathbf{v}^+$ a velocidade da partícula no primeiro e segundo
momentos de intercepção, e $x$ e $x^+$ os respectivos pontos onde ocorrem.
Denote-se ainda por $\varphi$ e $\varphi^+$ os ângulos que os
vectores $-\mathbf{v}$ e $\mathbf{v}^+$ fazem
com o vector normal à secção da curva $\partial(\text{conv}B)$ (direccionado para fora) entre os pontos $x$ e $x^+$.
Serão positivos se forem definidos no sentido anti-horário a partir
do vector normal, e
negativos caso contrário. Com estas definições,
tanto  $\varphi$ como $\varphi^+$
tomam valores no intervalo $[-\pi/2,\pi/2]$.

\begin{figure}[!ht]
\begin{center}
\includegraphics[width=0.45\columnwidth]{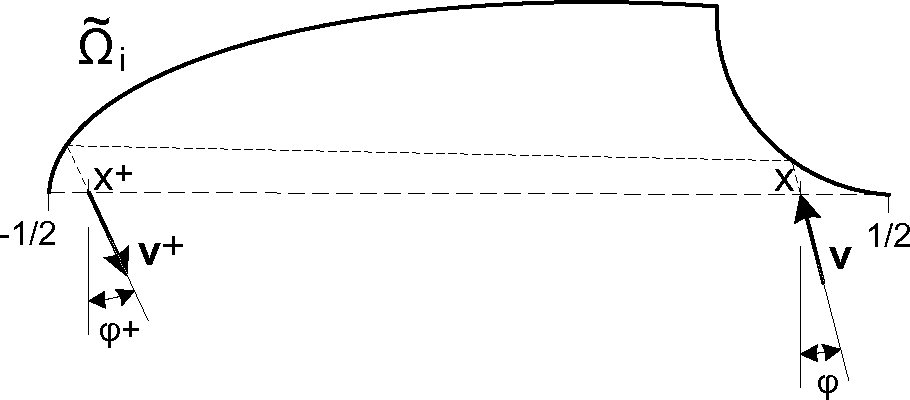}
\caption{Exemplo de trajectória que uma partícula descreve no interior de uma cavidade normalizada.}
\label{fig:cavidadeOmega}
\end{center}
\end{figure}

Representando as cavidades que caracterizam o contorno de $B$ por
subconjuntos $\Omega_1,\Omega_2, \ldots$, que no seu todo formam o conjunto
$\text{conv}B \setminus B$,
a resistência normalizada do corpo $B$ (equação~\eqref{eq:normalizacao}) assume a seguinte forma (cf. \cite{Plakhov07:CM, Plakhov07}):
\begin{equation}
\label{eq:RB2}
R(B)
=\frac{|\partial(\text{conv}B)|}{|\partial C_r|} \left(\frac{L_0}{L}+\sum_{i \ne 0}{\frac{L_i}{L}R(\tilde{\Omega}_i)}\right)
\text{,}
\end{equation}
sendo $L_0=|\partial(\text{conv}B)\cap \partial B|$ o comprimento da parte convexa
do contorno $\partial B$,
$L_i=|\partial(\text{conv}B)\cap {\Omega}_i|$, com $i=1,2,\ldots$, o tamanho da abertura da cavidade ${\Omega}_i$, 
e $R(\tilde{\Omega}_i)$ a resistência normalizada da cavidade $\tilde{\Omega}_i$, em relação a um segmento liso de tamanho unitário, com
\begin{equation}
\label{eq:R}
R(\tilde{\Omega}_i)=\frac{3}{8} \int_{-1/2}^{1/2}\int_{-\pi/2}^{\pi/2} \left(
1+\cos\left( \varphi^+(x,\varphi) -\varphi \right) 
\right) \cos \varphi\, \dd\varphi\,\dd x
\text{.}
\end{equation}
A função $\varphi^+$ deve ser vista como o ângulo de saída
duma partícula que interage com uma cavidade $\tilde{\Omega}_i$ de abertura de tamanho
unitário, centrada no eixo dos $x$ (ver ilustração da Figura~\ref{fig:cavidadeOmega}),
e com a forma de $\Omega_i$ --- relacionando-se com esta por um factor de proporcionalidade $1/L_i$.

Da equação~\eqref{eq:RB2} percebemos que a resistência de um corpo $B$
pode ser vista como uma média ponderada ($\sum_i L_i/L=1$) das resistências das
cavidades individuais que caracterizam toda a sua fronteira (incluindo
``cavidades'' lisas), multiplicada por um factor que relaciona os
perímetros dos corpos $\text{conv}B$ e $C_r$.
Assim, maximizarmos a resistência do corpo $B$ equivale
a maximizarmos o perímetro de $\text{conv}B$ ($|\partial(\text{conv}B)|\le |\partial C_r|$)
e as resistências individuais das cavidades $\Omega_i$.

Encontrada a forma óptima $\Omega^*$, que maximize a funcional~\eqref{eq:R},
o corpo $B$ de resistência máxima será aquele cuja
fronteira seja formada unicamente pela concatenação de pequenas cavidades com essa forma.
Podemos então restringir o nosso problema à subclasse de corpos $B$ que tenham
a sua fronteira integralmente preenchida por cavidades iguais, e com isso
admitir, sem qualquer perda de generalidade, que
cada cavidade $\Omega_i$ ocupa o lugar de um arco de círculo de tamanho $\varepsilon\ll r$.
Como $L_i=2r\sin(\varepsilon/2r)$, a razão entre os perímetros assume o valor
\begin{equation}
\label{eq:RxPerimetros}
\frac{|\partial(\text{conv}B)|}{|\partial C_r|}
=\frac{ \sin(\varepsilon/2r)}{\varepsilon/2r}
\approx 1- \frac{(\varepsilon/r)^2}{24}\text{,}
\end{equation}
ou seja, dado um corpo $B$ de fronteira formada por
cavidades com a forma $\Omega$, de~\eqref{eq:RB2} e~\eqref{eq:RxPerimetros}
concluímos que a resistência total do corpo
será igual à resistência da cavidade individual $\Omega$,
menos uma pequena fracção desse valor,
que será negligenciável quando $\varepsilon\ll r$,
\begin{equation}
\label{eq:Raprox}
R(B)\approx R(\Omega)-\frac{(\varepsilon/r)^2}{24}R(\Omega)\text{.}
\end{equation}

Assim, as nossas pesquisas têm como objectivo encontrar formas de cavidades $\Omega$ que maximizem o valor da funcional~\eqref{eq:R}, cujo supremo sabemos situar-se
no intervalo
\begin{equation}
\label{eq:RminMax}
1\le \text{sup}_\Omega R(\Omega) \le 1.5
\text{,}
\end{equation}
como facilmente se comprova usando~\eqref{eq:R}:
se $\Omega$ for um segmento liso, $\varphi^+(x,\varphi)=-\varphi$ e
$R(\Omega)=\frac{3}{8} \int_{-1/2}^{1/2}\int_{-\pi/2}^{\pi/2} \left(
1+\cos\left(  2\varphi \right) 
\right) \cos \varphi\, \dd\varphi\,\dd x=1
$;
nas condições de resistência máxima $\varphi^+(x,\varphi)=\varphi$,
logo $
R(\Omega)\le\frac{3}{8} \int_{-1/2}^{1/2}\int_{-\pi/2}^{\pi/2} 2 \cos \varphi\, \dd\varphi\,\dd x
=1.5$.

\section{Estudo numérico do problema}
\label{sec:estNumer}

Na classe de problemas que estamos a considerar, apenas para algumas formas $\Omega$ muito elementares é possível desvendar a fórmula analítica da sua resistência~\eqref{eq:R}, como são disso exemplo as formas rectangulares e triangulares anteriormente referenciadas.
Para formas um pouco mais elaboradas, o cálculo analítico torna-se rapidamente
demasiado complexo, senão impossível, dada a grande dificuldade em conhecermos a função $\varphi^+: [-1/2,1/2]\times[-\pi/2,\pi/2]\rightarrow[-\pi/2,\pi/2]$,
que, como sabemos, está intimamente relacionada com o formato da cavidade $\Omega$.
Assim, o recurso à computação numérica surge como a abordagem natural
e inevitável para se poder investigar essa classe de problemas.

Desenvolveram-se modelos computacionais que simulam a dinâmica de bilhar no interior
de cada uma das formas $\Omega$ estudadas. Os algoritmos de construção desses modelos,
bem como os responsáveis pelo cálculo numérico da resistência associada,
foram implementados usando a linguagem de programação C,
dado o esforço computacional envolvido (a linguagem C foi criada em 1972 por Dennis Ritchie;
para o seu estudo sugerimos, entre a extensa documentação disponível, aquele que é o livro de referência da linguagem, da autoria de Brian Kernighan e do próprio Dennis Ritchie, \cite{LangC}). A eficiência do código objecto,
gerado pelos compiladores de C, permitiu que a aproximação numérica de~\eqref{eq:R} fosse realizada com um número suficientemente elevado de subdivisões dos intervalos de integração --- entre algumas centenas e vários milhares (até $5000$). Os resultados foram, por isso, obtidos com uma precisão que atingiu em alguns dos casos $10^{-6}$. Essa precisão foi controlada por observação da diferença entre aproximações sucessivas da resistência $R$ que se iam obtendo com o aumento do número de subdivisões.

Para a maximização da resistência dos modelos idealizados,
usaram-se os algoritmos de optimização global da
\textit{toolbox} ``\emph{Genetic Algorithm and Direct Search}''
(versão 2.0.1 (R2006a), documentada em \cite{toolbox}),
uma colecção de funções que estende as capacidades
de optimização do sistema de computação numérica MATLAB. 
A opção pelos métodos Genéticos e de procura Directa deveu-se essencialmente ao facto de os mesmos
não requererem qualquer informação acerca do gradiente da função objectivo nem de derivadas de ordem superior --- como a forma analítica da função resistência é em geral desconhecida (dado que depende de $\varphi^+(x,\varphi)$), esse tipo de informação, caso fosse necessária, teria que ser obtida por aproximação numérica, algo que dificultaria imenso o processo de optimização.
O sistema de computação MATLAB (versão 7.2 (R2006a)) foi também escolhido por
dispor de funcionalidades que permitiram que se usasse para 
função objectivo a subrotina compilada em C de cálculo da resistência, bem como
a função $\varphi^+(x,\varphi)$ por si invocada.



\subsection{``Dupla Parábola'': uma forma bidimensional que maximiza a resistência}
\label{sec:DuplaParab}

No estudo numérico que os autores realizaram em \cite{Plakhov07:CM, Plakhov07} procuraram-se
formas $\Omega_f$ definidas por funções $f:[-1/2,1/2]\rightarrow \mathbb{R}_+$ contínuas e seccionalmente diferenciáveis: 
\begin{equation}
\label{eq:Omegaf}
\Omega_f=\left\{(x,y):\,-1/2\le x \le 1/2,\; 0\le y \le f(x)\right\}\text{.}
\end{equation}
Iniciou-se a procura da resistência máxima na classe de funções contínuas $f$ com derivada $f'$ seccionalmente constante, alargando-se depois o estudo a classes de funções com a segunda derivada $f''$ seccionalmente constante. No primeiro dos casos o contorno de $\Omega_f$
é uma linha poligonal, e no segundo, uma curva composta por arcos de parábolas.
Não se tendo conseguido com as formas $\Omega_f$ superar o valor de resistência $R=1.44772$,
decidimos, neste novo estudo, estender a procura a formas diferentes das consideradas
em~\eqref{eq:Omegaf}. Estudámos formas $\Omega^g$ definidas por funções de $y$ em $x$ do seguinte modo:
\begin{equation}
\label{eq:Omegag}
\Omega^g=\left\{(x,y):\,0\le y \le h,\; -g(y)\le x \le g(y)\right\}\text{,}
\end{equation}
onde $h>0$ e $g:[0,h] \rightarrow \mathbb{R}_0^+$ é uma função contínua
com $g(0)=1/2$ e $g(h)=0$. 

O novo problema de resistência máxima por nós estudado pode então ser formulado da seguinte forma:
\begin{quotation}{\sl
Encontrar $\sup_g R(\Omega^g)$ nas
funções $g:[0,h] \rightarrow \mathbb{R}_0^+$ contínuas e seccionalmente diferenciáveis,
tais que $g(0)=1/2$ e $g(h)=0$, com $h>0$.
}\end{quotation}

À semelhança do estudo que se fez para os conjuntos $\Omega_f$,
na procura das formas $\Omega^g$ consideraram-se funções $g$ seccionalmente lineares e seccionalmente quadráticas. Se na classes das funções lineares não se conseguiu qualquer ganho
de resistência relativamente aos resultados obtidos para os conjuntos $\Omega_f$, já
nas funções quadráticas os resultados superaram as melhores expectativas: encontrou-se uma forma de cavidade $\Omega^g$ que apresenta uma resistência $R=1.4965$, um valor já muito próximo do seu limite teórico de $1.5$. Este é seguramente um resultado muito interessante.
Efectuaram-se ainda alguns testes com funções polinomiais de ordem superior
ou descritas por secções cónicas específicas
mas, não se tendo verificado qualquer ganho adicional na maximização da resistência,
optou-se por não se reportar os respectivos resultados.
Segue-se então a descrição do melhor resultado que se obteve, encontrado na classe das funções $x=\pm g(y)$ quadráticas.

\begin{figure}[!ht]
\begin{center}
\begin{tabular}{c c}
\includegraphics*[height=0.37\columnwidth]{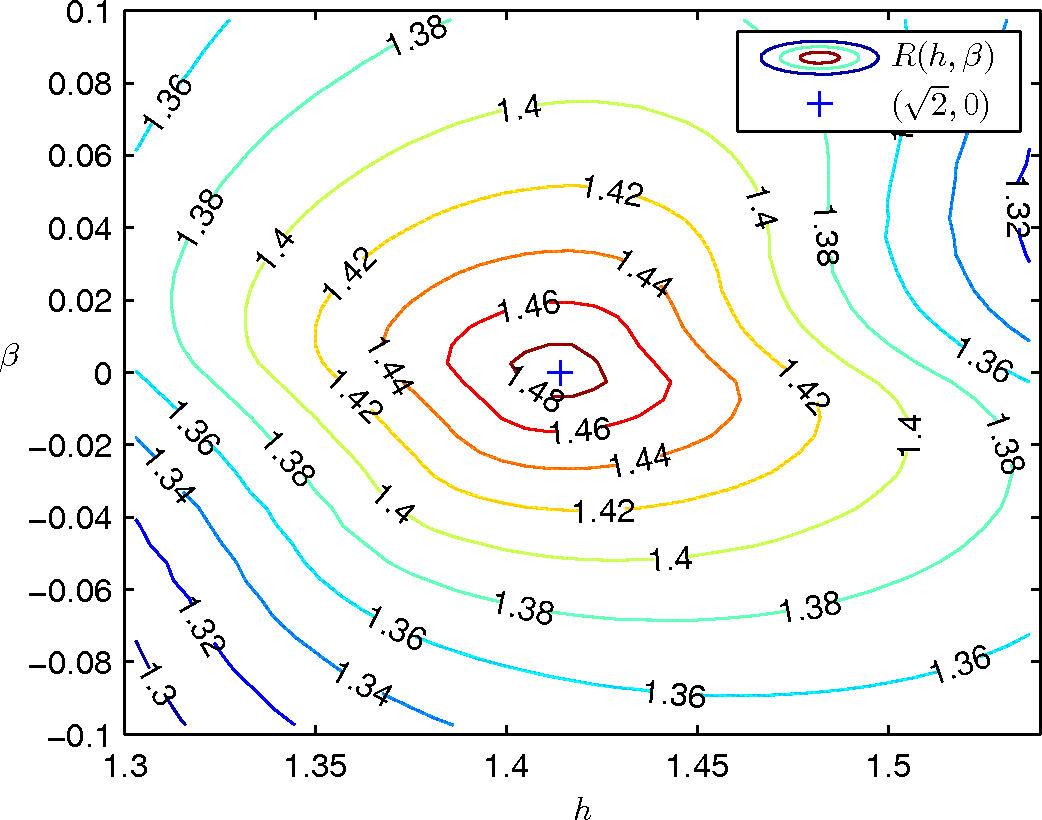}
&
\includegraphics*[height=0.37\columnwidth]{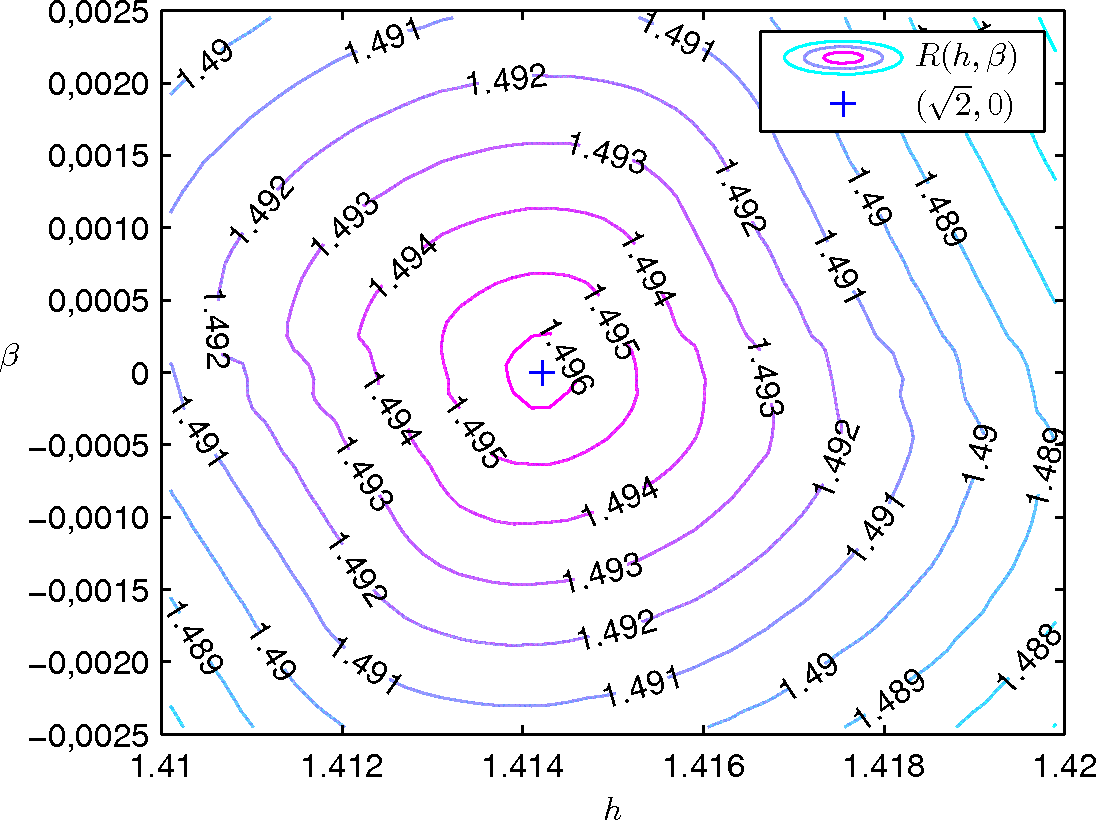} \\
(a)&(b)\\
\end{tabular}
\caption{Curvas de nível da função $R(h,\beta)$.}
\label{fig:curvNivel}
\end{center}
\end{figure}

\begin{figure}[!ht]
\begin{center}
\includegraphics*[width=0.5\columnwidth]{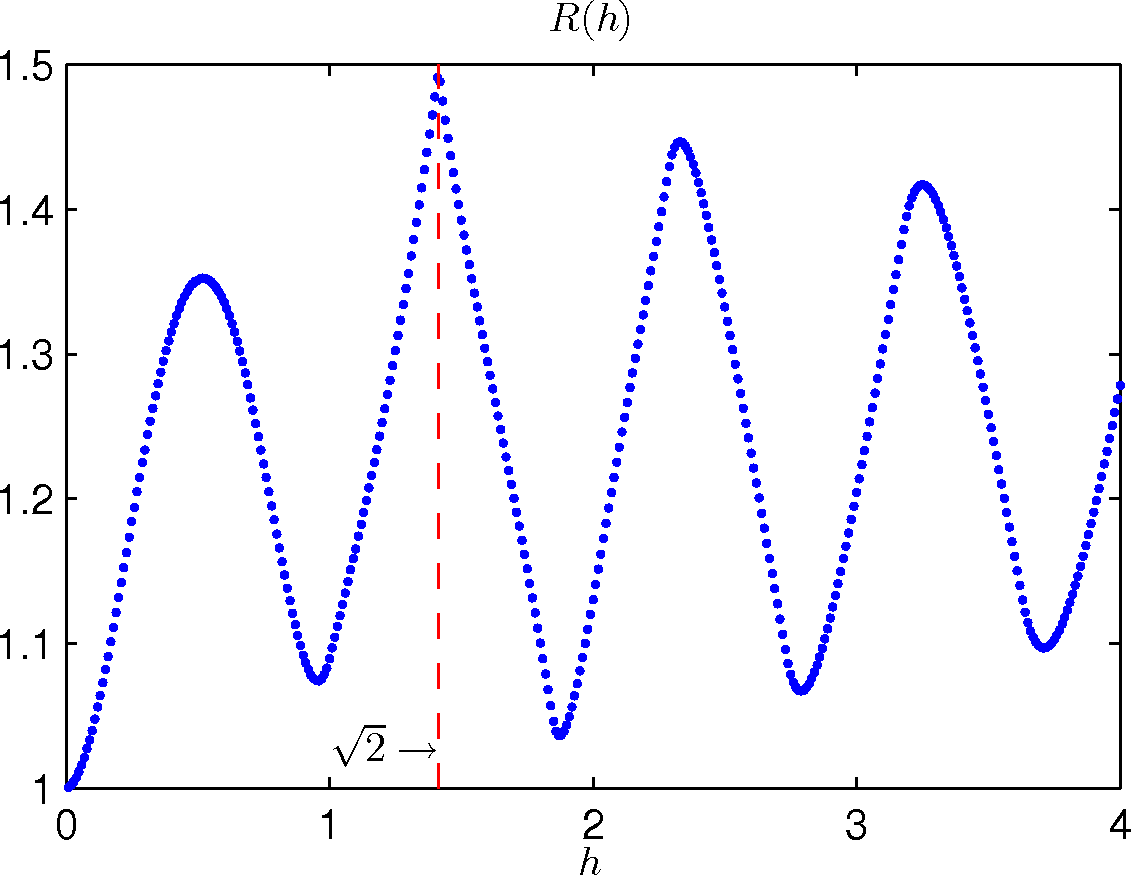}
\caption{Gráfico da resistência $R(h)$ para $\beta=0$.}
\label{fig:Res_h}
\end{center}
\end{figure}

Estudou-se o valor da resistência de conjuntos $\Omega^g$, tal como definidos em~\eqref{eq:Omegag}, na classe de funções quadráticas,
$$
g_{h,\beta}(y) = 
 \alpha y^2 + \beta y +1/2, \text{ para } 0 \le y \le h\,,
$$
onde $h>0$ e $\alpha= \frac{-\beta h -1/2}{h^2}$ (dado que $g_{h,\beta}(h) =0$). Na optimização da curva, fizeram-se variar os dois parâmetros de configuração da função:
$h$, a altura da curva $\partial\Omega^g$, e $\beta$, o seu declive na origem ($g'(0)$).
Nesta classe de funções os algoritmos de optimização convergiram rapidamente para um resultado muito interessante:
a resistência máxima foi atingida com $h=1.4142$ e $\beta=0.0000$, e assumiu
o valor $R=1.4965$, ou seja, um valor $49.65\%$ acima da resistência da forma rectilínea.
É efectivamente um resultado muito importante:
\begin{enumerate}
\item representa um ganho considerável no valor da resistência,
relativamente ao melhor resultado anterior (obtido em \cite{Plakhov07:CM, Plakhov07}), 
que se situava $44.77\%$ acima do valor de referência;
\item o conjunto $\Omega^g$ correspondente tem uma forma bastante mais simples do que a do conjunto $\Omega_f$ associada ao melhor resultado anterior, uma vez que
é formada por dois arcos de parábolas simétricos, quando a anterior
era constituída por catorze desses arcos;
\item esse novo valor de resistência está já muito próximo do seu limite máximo teórico
que, como se sabe, situa-se $50\%$ acima do valor de referência;
\item os parâmetros óptimos parecem assumir valores que dão ao conjunto $\Omega^g$
uma configuração com características muito especiais, como a seguir se perceberá.
\end{enumerate}
Repare-se que os parâmetros óptimos parecem aproximar-se
dos valores $h=\sqrt{2}=1.41421\ldots$ e $\beta=0$.
Coloca-se então a seguinte questão:
\begin{quotation}{\sl
Não serão esses os valores exactos dos parâmetros óptimos?
}\end{quotation}
A representação gráfica da função $R(h,\beta)$ através de curvas
de nível, Figura~\ref{fig:curvNivel}, está em perfeita concordância com
essa possibilidade --- repare-se que as curvas de nível parecem 
perfeitamente centradas no ponto de coordenadas $(\sqrt{2},0)$,
assinalado na figura por um ``$+$''.
Veja-se também, na Figura~\ref{fig:Res_h}, o gráfico da resistência $R(h)$ para $\beta=0$,
onde é igualmente perceptível uma surpreendente elevação da resistência quando $h\rightarrow\sqrt{2}$.
Calculou-se então numericamente a resistência da cavidade $\Omega^{g_{h\beta}}$
com os valores exactos $h=\sqrt{2}$ e $\beta=0$, tendo o resultado
confirmado o valor $1.49650$.

\begin{figure}[!ht]
\begin{center}
\begin{tabular}{c c c}
\includegraphics*[height=0.35\columnwidth]{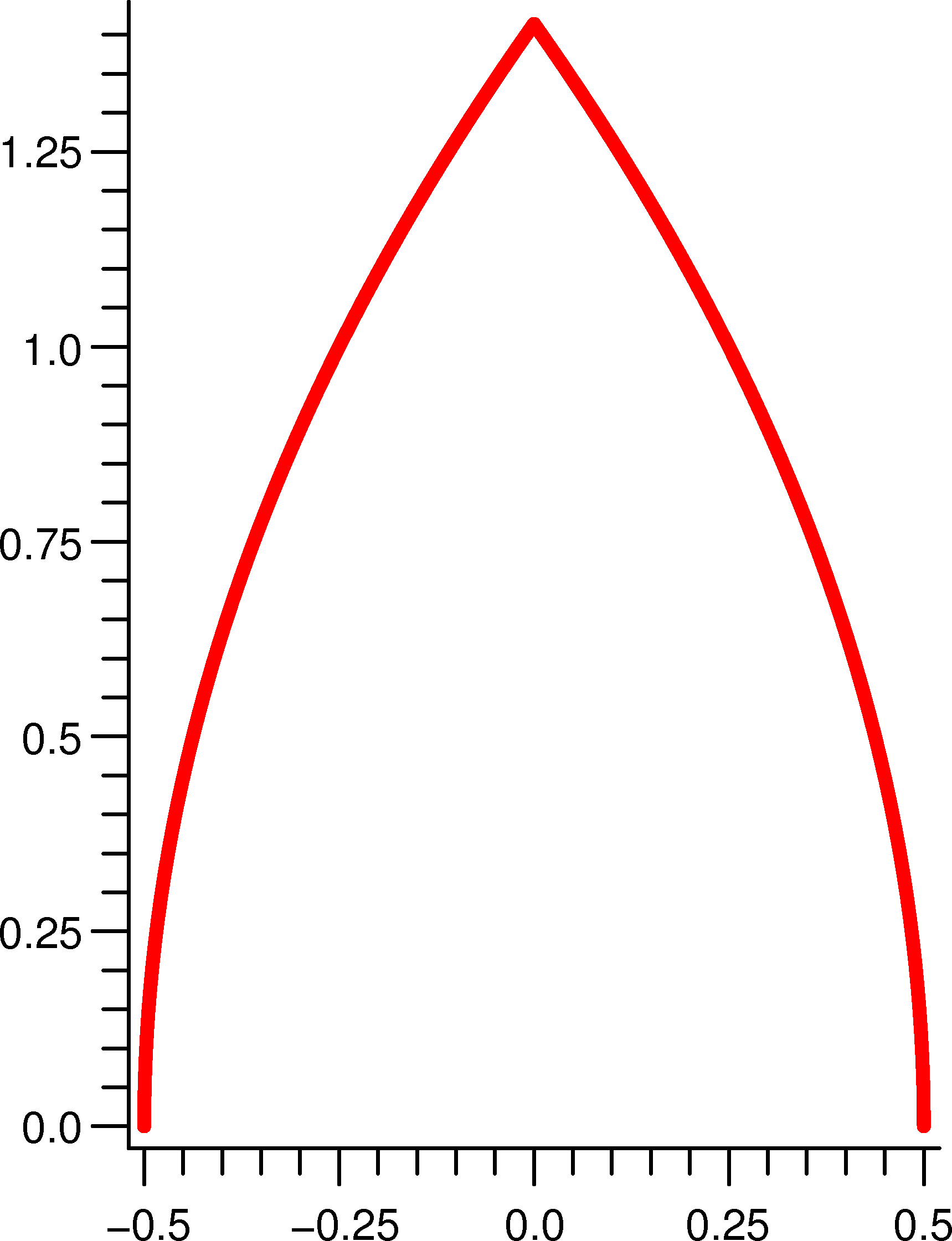}
&
&
\includegraphics*[height=0.35\columnwidth]{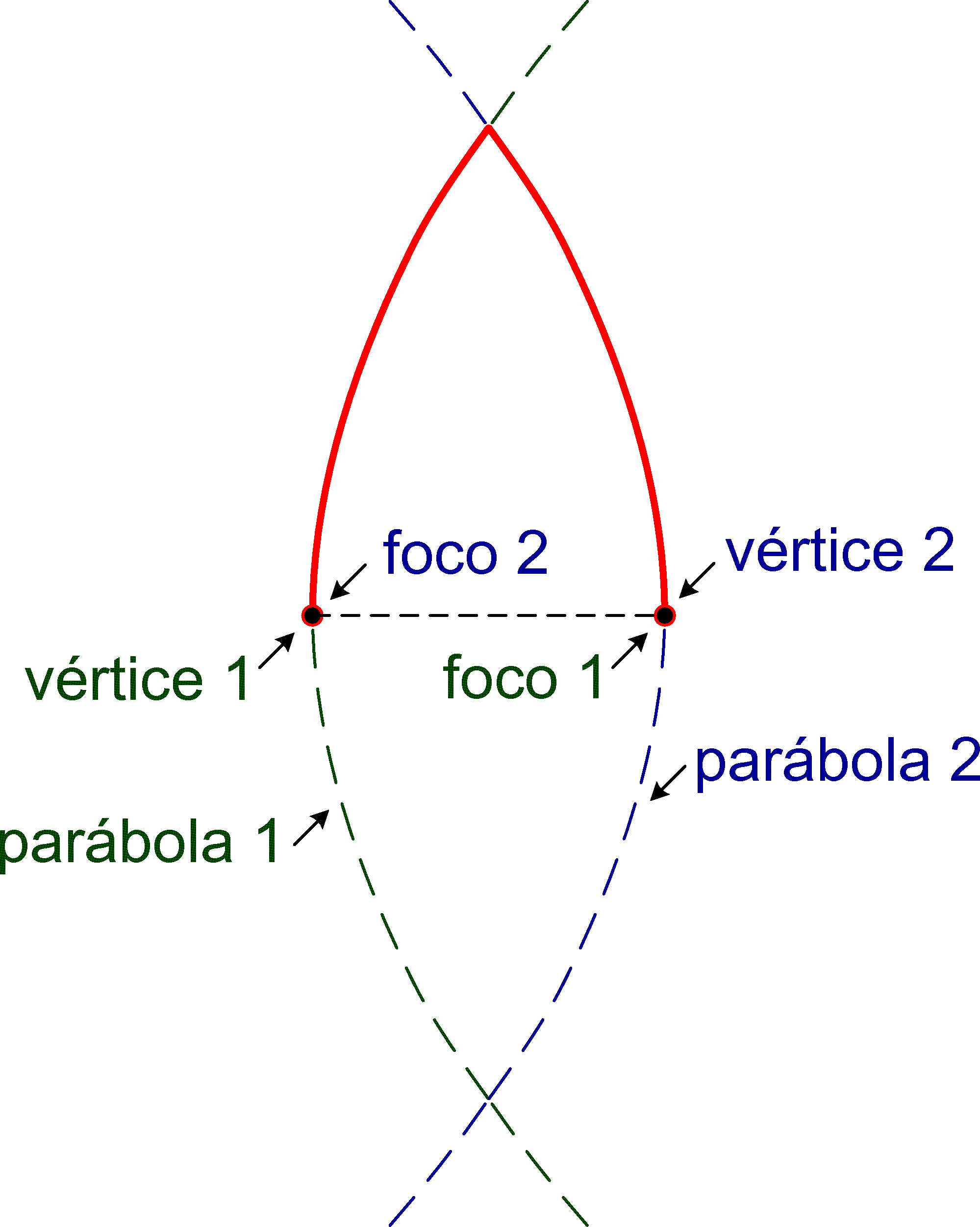} \\
(a)&&(b)\\
\end{tabular}
\caption{Forma 2D (quase) óptima --- futuramente conhecida por \emph{Dupla Parábola}.}
\label{fig:parabOpt}
\end{center}
\end{figure}

Existe ainda uma razão adicional que sugere também uma resposta afirmativa à
questão formulada.
A forma do conjunto $\Omega^{g_{h,\beta}}$ com $h=\sqrt{2}$ e $\beta=0$ 
é um caso particular a que estão associadas características especiais
que poderão justificar o elevado valor de resistência que apresenta.
As duas secções da forma são arcos equivalentes de duas parábolas
de eixos horizontais e concavidades voltadas uma para a outra --- ver Figura~\ref{fig:parabOpt}.
Mas a particularidade da configuração reside no facto do eixo das parábolas
coincidir com a linha de entrada da cavidade (eixo dos $x$), e o foco de cada uma situar-se no vértice da outra.
Repare-se que, sendo a equação da parábola do lado direito dada por
$x=g_{\sqrt{2},0}(y)$, obtém-se $y^2=-4(x-1/2)$, confirmando
que o vértice é em $(1/2,0)$ e que o foco dista deste uma unidade. 
Desse modo, por razões de simetria,
os focos e os vértices das duas parábolas situam-se nos
extremos da abertura da cavidade, $(-1/2,0)$ e $(1/2,0)$,
tal como se ilustra no esquema da Figura~\ref{fig:parabOpt}b.

\begin{figure}[!ht]
\begin{center}
\includegraphics*[width=0.6\columnwidth]{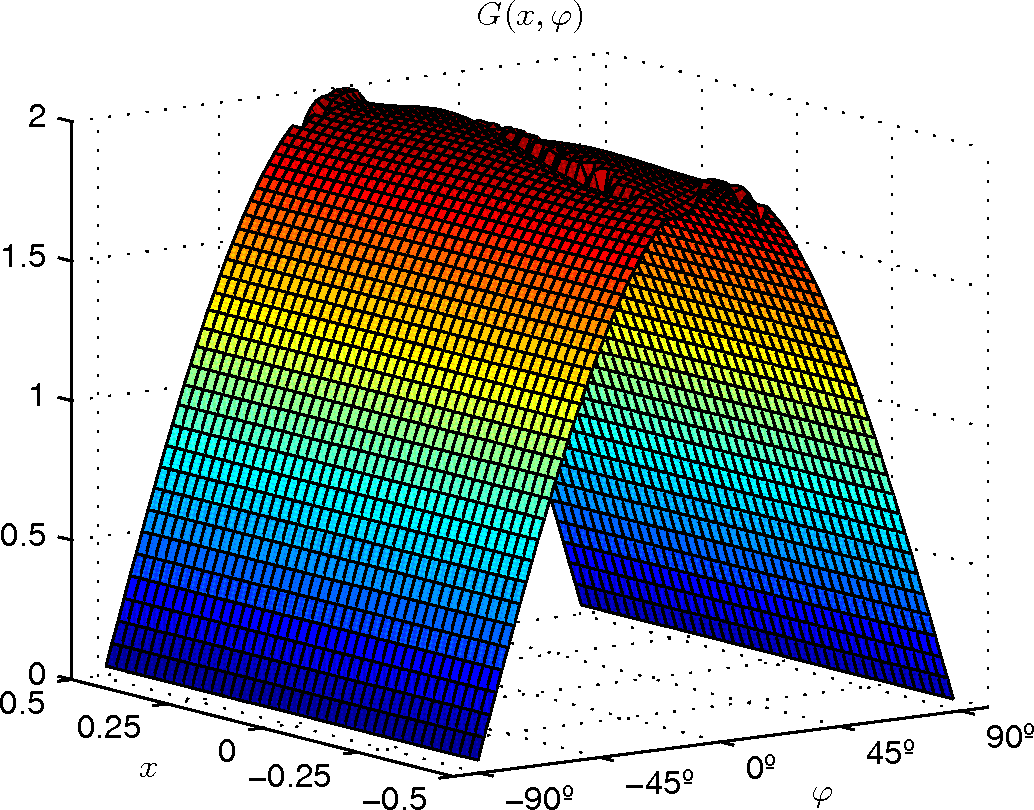}
\caption{Gráfico da função integranda $G(x,\varphi)=\left(
1+\cos\left( \varphi^+(x,\varphi) -\varphi \right) 
\right) \cos \varphi$.}
\label{fig:funInteg}
\end{center}
\end{figure}

O elevado valor da resistência desta forma de cavidade está ainda expresso,
embora de modo implícito, no gráfico da função integranda da funcional~\eqref{eq:R},
apresentado na Figura~\ref{fig:funInteg}. Trata-se de uma superfície com uma
forma bastante suave em quase todo o seu domínio e, mais importante, quase não varia
com $x$ e parece descrever a forma da função $2\cos\varphi$
na direcção do eixo dos $\varphi$. Estes atributos, quando verificados
na sua plenitude, estão associados ao maior valor de resistência admissível,
$R=1.5$.\footnote{Se a função integranda tomar a forma $G(x,\varphi)=2\cos\varphi$,
de~\eqref{eq:R} obtém-se $R=1.5$.}

Esta forma de cavidade parece tratar-se efectivamente de um caso muito particular. Contrariamente ao que se passou com todas as outras formas anteriormente estudadas, a função integranda aparenta uma forma bastante suave, apresentando apenas pequenas irregularidades para ângulos $\varphi$ de pequena amplitude.
Atendendo a essa característica, e tendo em conta que $G(x,\varphi)$ quase não depende de $x$,
a resistência foi calculada, para esta forma em particular, usando a regra de Simpson $1/3$ na integração em ordem a $\varphi$. 
A dupla integração na equação~\eqref{eq:R} foi então aproximada numericamente
pela seguinte expressão:
\begin{equation}
\label{eq:RnsSimpson}
R=\frac{1}{2} \Delta x \Delta \varphi \sum_{i=N_x/2+1}^{N_x}\sum_{k=1}^{N_\varphi-1} w_k \left(
1+\cos\left( \varphi^+(x_i,\varphi_k) -\varphi_k \right) 
\right) \cos \varphi_k\text{,}
\end{equation}
com $w_k=2$ para $k$ ímpar e $w_k=1$ para $k$ par,
$x_i=- 1/2+(i-1/2) \Delta x$, $\Delta x=1/N_x$, $\varphi_k=-\pi/2+ k \Delta \varphi$ e  $\Delta \varphi=\pi/N_\varphi$. $N_x$ e $N_\varphi$
são o número de subintervalos a considerar na integração das variáveis $x$ e $\varphi$ (ambos números pares), 
respectivamente, e $\Delta x$ e $\Delta \varphi$ os incrementos para as correspondentes variáveis discretas. Dado que a forma $\Omega^{g_{\sqrt{2},0}}$ apresenta simetria horizontal, o primeiro somatório da expressão considera apenas a segunda metade do intervalo de integração da variável $x$.

Para que passe a ser facilmente referenciável, esta forma de cavidade (Figura~\ref{fig:parabOpt}a) passará, a partir de agora, a ser designada  simplesmente por ``\emph{Dupla Parábola}''. Assim, no contexto deste trabalho, o termo ``Dupla Parábola'' deve ser sempre entendido como o nome da cavidade cuja forma é descrita por duas parábolas que,
para além de serem geometricamente iguais, encontram-se ``encaixadas'' na disposição peculiar
que referimos.

Uma vez que a resistência da Dupla Parábola
assume um valor já muito próximo do seu limite teórico, numa derradeira
tentativa de se vir a conseguir atingir esse limite, resolveu-se estender
ainda o estudo a outras classes de funções $g(y)$ que admitissem a Dupla Parábola
como caso particular ou que permitissem configurações próximas dessa forma
quase óptima. Em todos esses casos os melhores resultados foram invariavelmente obtidos
quando a forma das curvas se aproximou da forma da Dupla Parábola, sem nunca terem superado o valor $R=1.4965$.
Começou-se por considerar funções $g(y)$ seccionalmente quadráticas,
incluindo curvas \textit{splines}, sem que se conseguissem resultados
interessantes; apenas para funções $g(y)$ de $2$ ou $3$ segmentos foi possível
aproximarmo-nos da resistência e da forma da Dupla Parábola.
Consideraram-se também funções $g(y)$ cúbicas e biquadráticas\footnote{Nas curvas biquadráticas, o ponto de intercepção da trajectória da partícula com a fronteira da cavidade
é calculado resolvendo uma equação do
4º grau. As raízes dessa equação foram obtidas numericamente usando
o método descrito em \cite{Hook90}. As equações de ordem inferior foram sempre resolvidas recorrendo às conhecidas fórmulas resolventes.}, mas em ambos os casos o processo de optimização aproximou-as a curvas de ordem quadrática, com os coeficientes de maior ordem a tomarem
valores quase nulos.
Estudou-se o problema na classe das secções cónicas, considerando-se, para faces laterais da cavidade, dois arcos simétricos quer de uma elipse quer de uma hipérbole.
Também nestes casos os arcos assumiram uma forma muito próxima dos arcos de parábolas.

Sendo a Dupla Parábola a melhor forma encontrada, e tratando-se de uma forma quase óptima,
na secção que se segue é objecto de um estudo aprofundado, de natureza essencialmente analítica, onde se tentam perceber as razões do seu bom desempenho.

\section{Caracterização das reflexões na forma ``Dupla Parábola''}
\label{sec:caract}

\begin{figure}[!hb]
\begin{center}
\begin{tabular}{c c c}
\includegraphics*[width=0.2\columnwidth]{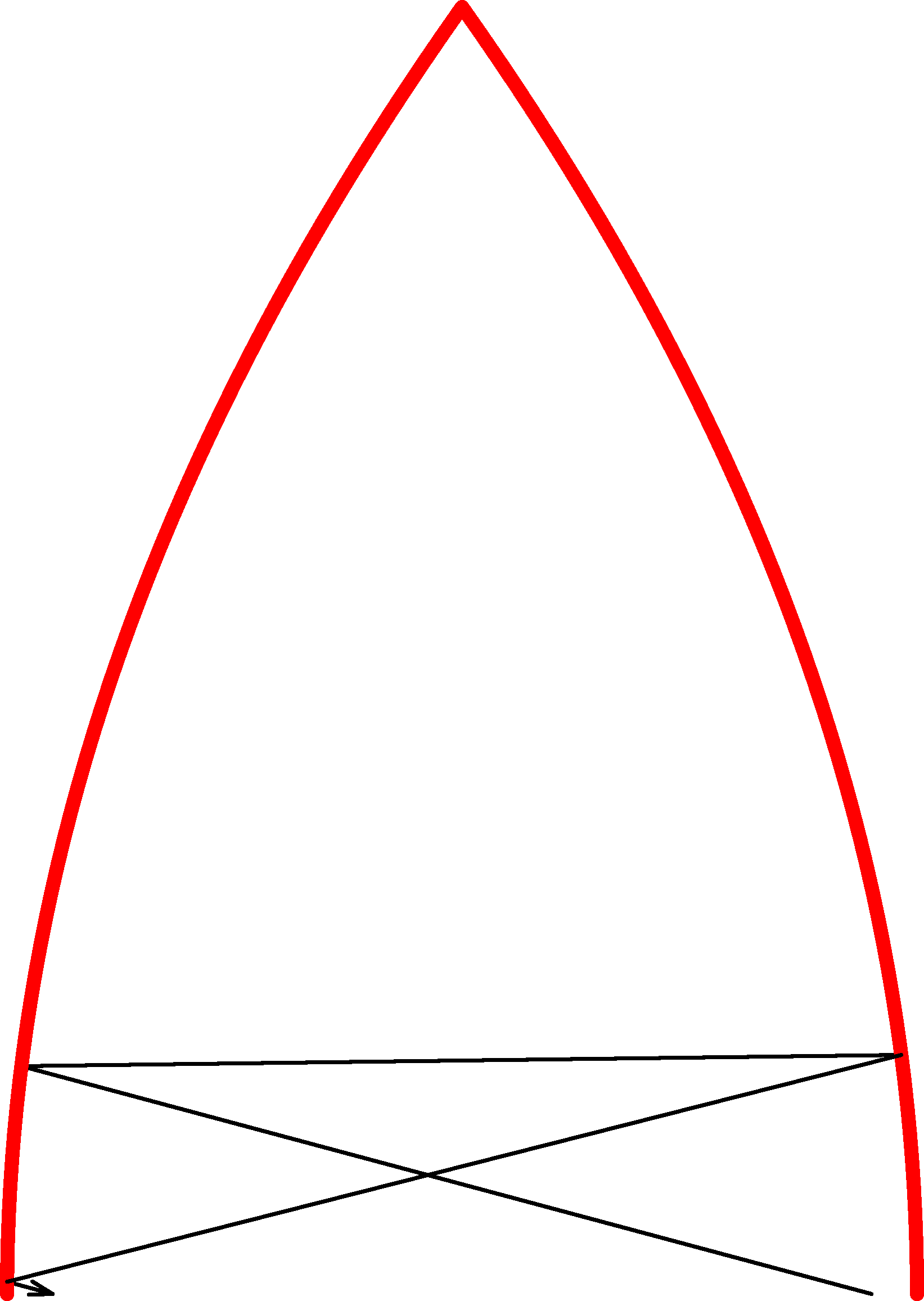} & \includegraphics*[width=0.2\columnwidth]{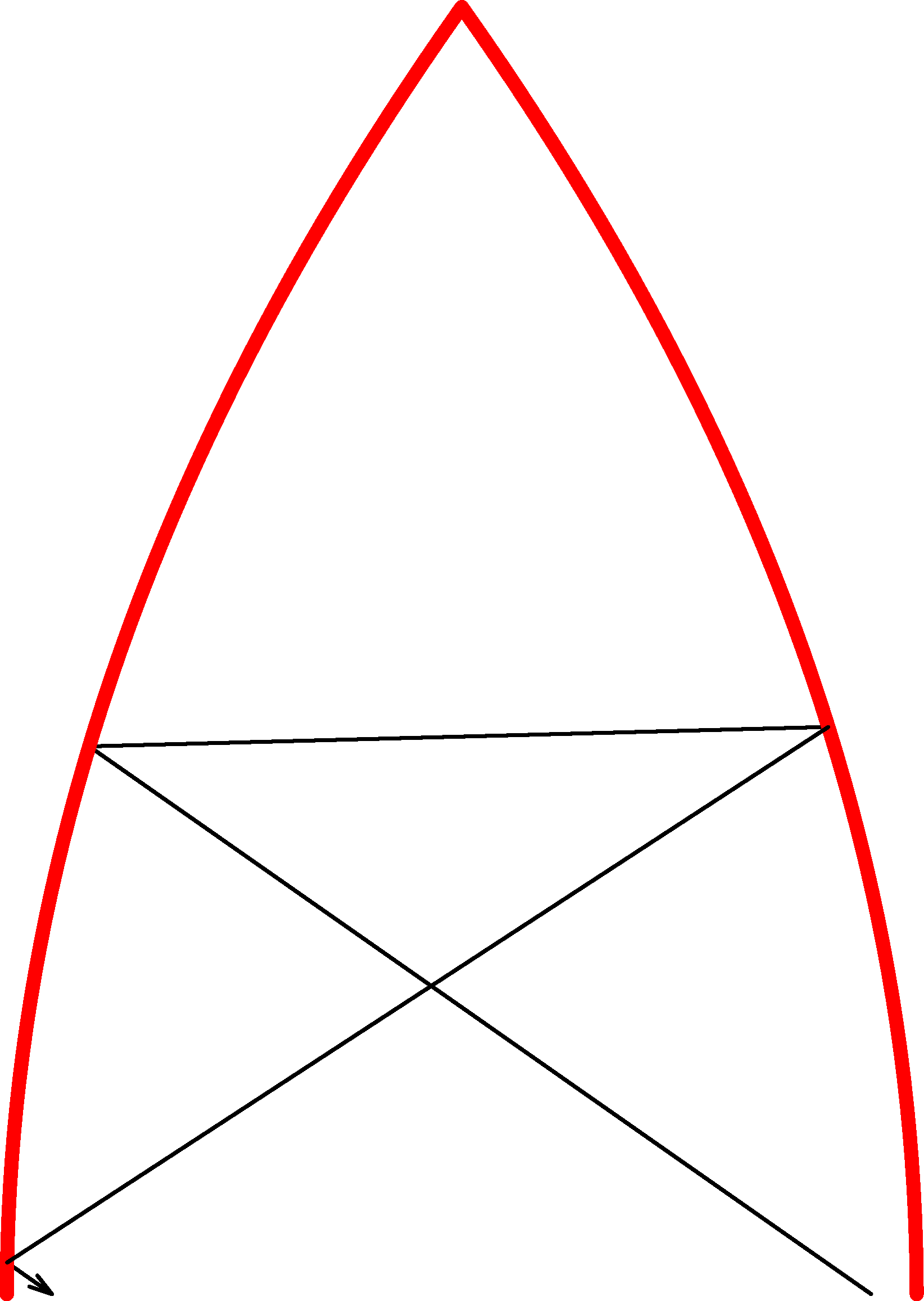} &
\includegraphics*[width=0.2\columnwidth]{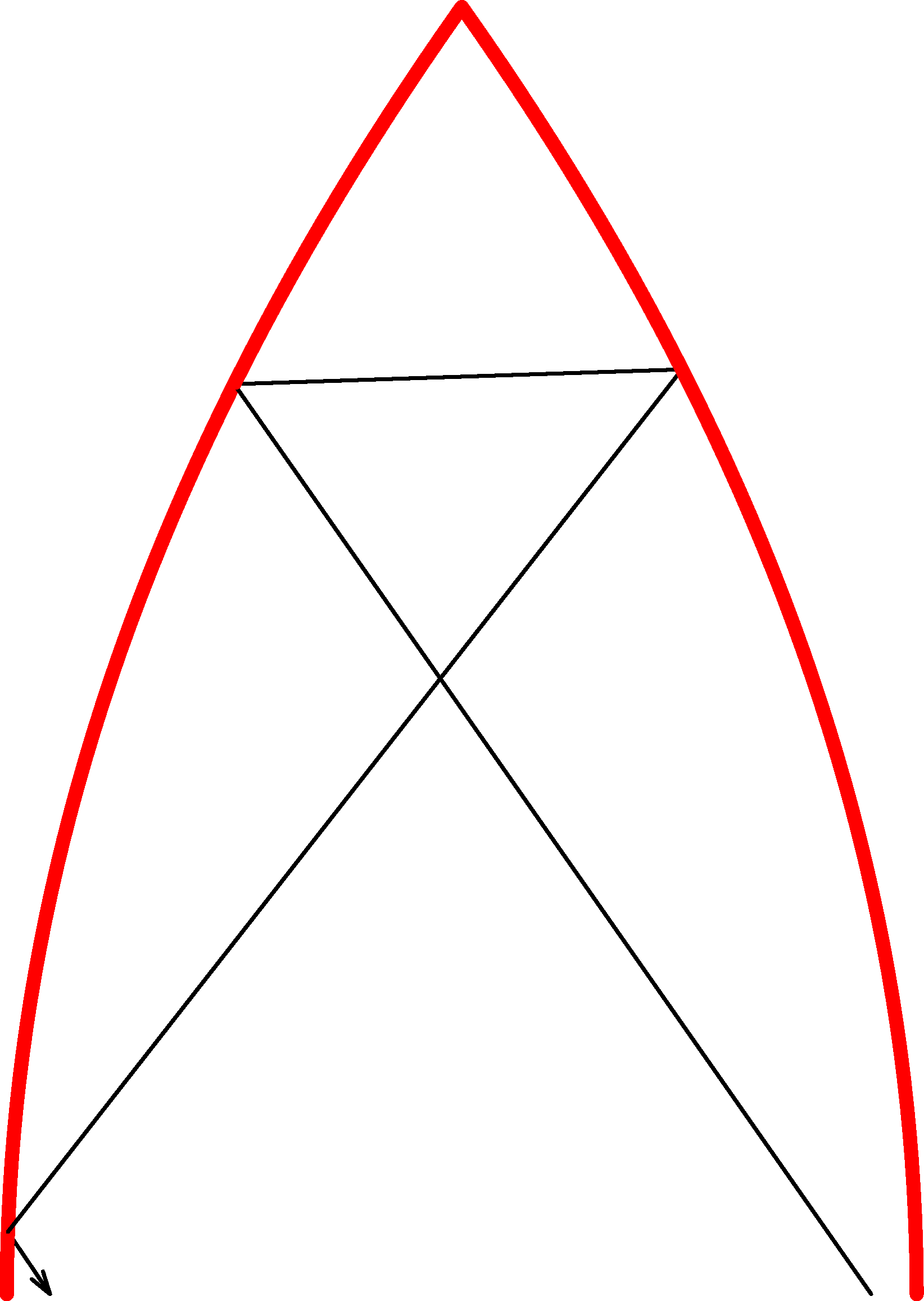} \\
(a) $x=0.45$, $\varphi=75^\circ$.&(b) $x=0.45$, $\varphi=55^\circ$.&(c) $x=0.45$, $\varphi=35^\circ$.\\
\includegraphics*[width=0.2\columnwidth]{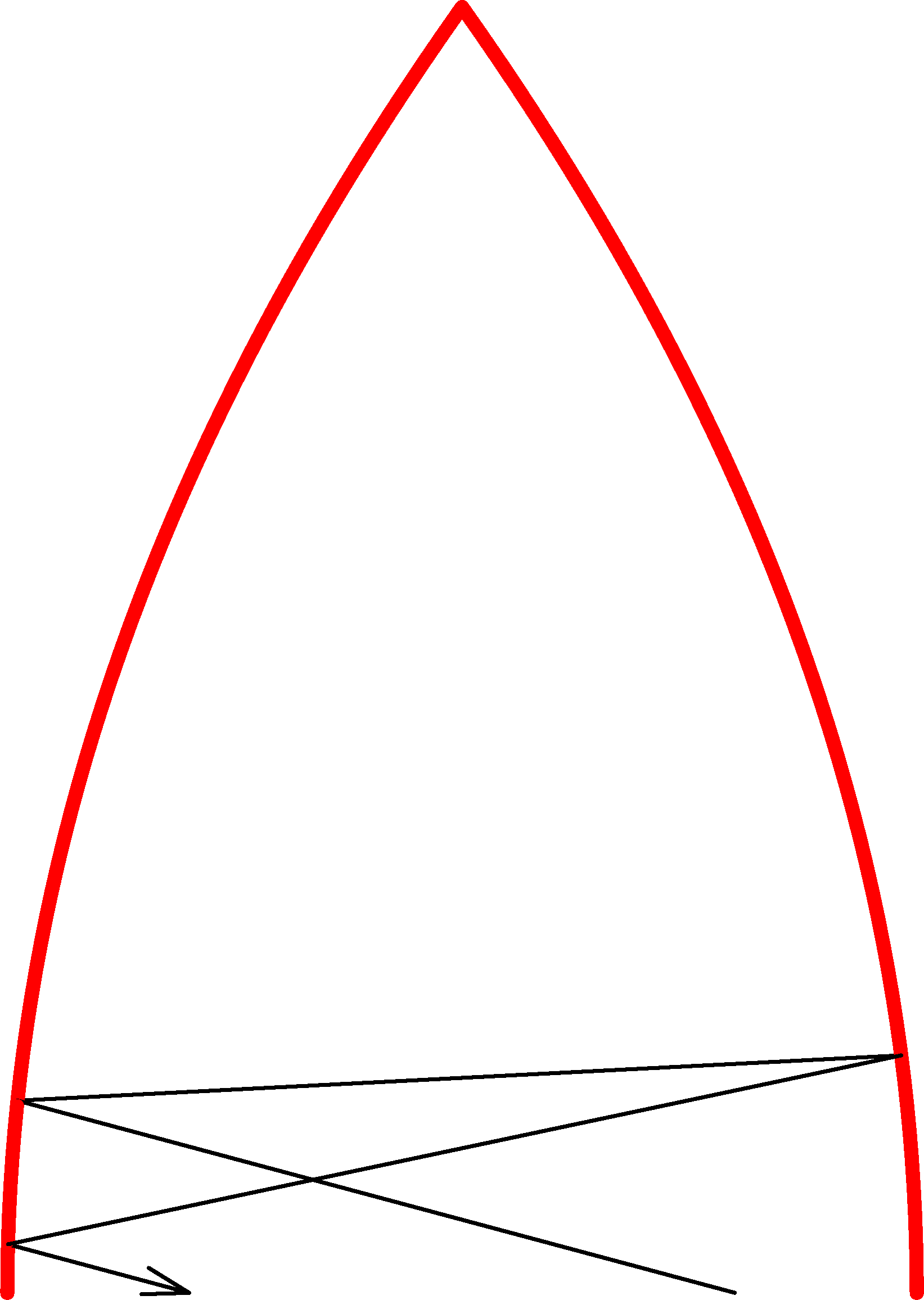} &
\includegraphics*[width=0.2\columnwidth]{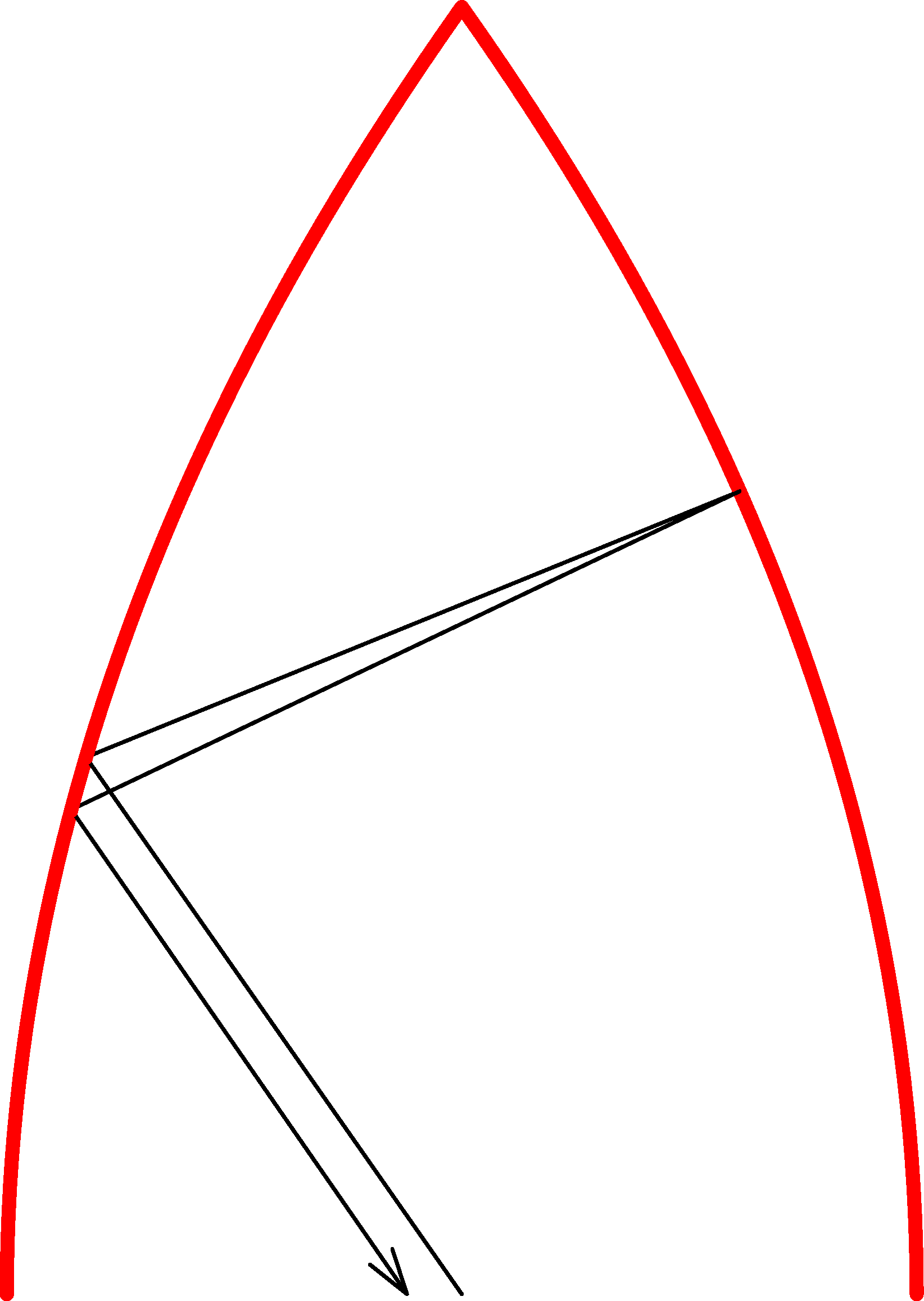} &
\includegraphics*[width=0.2\columnwidth]{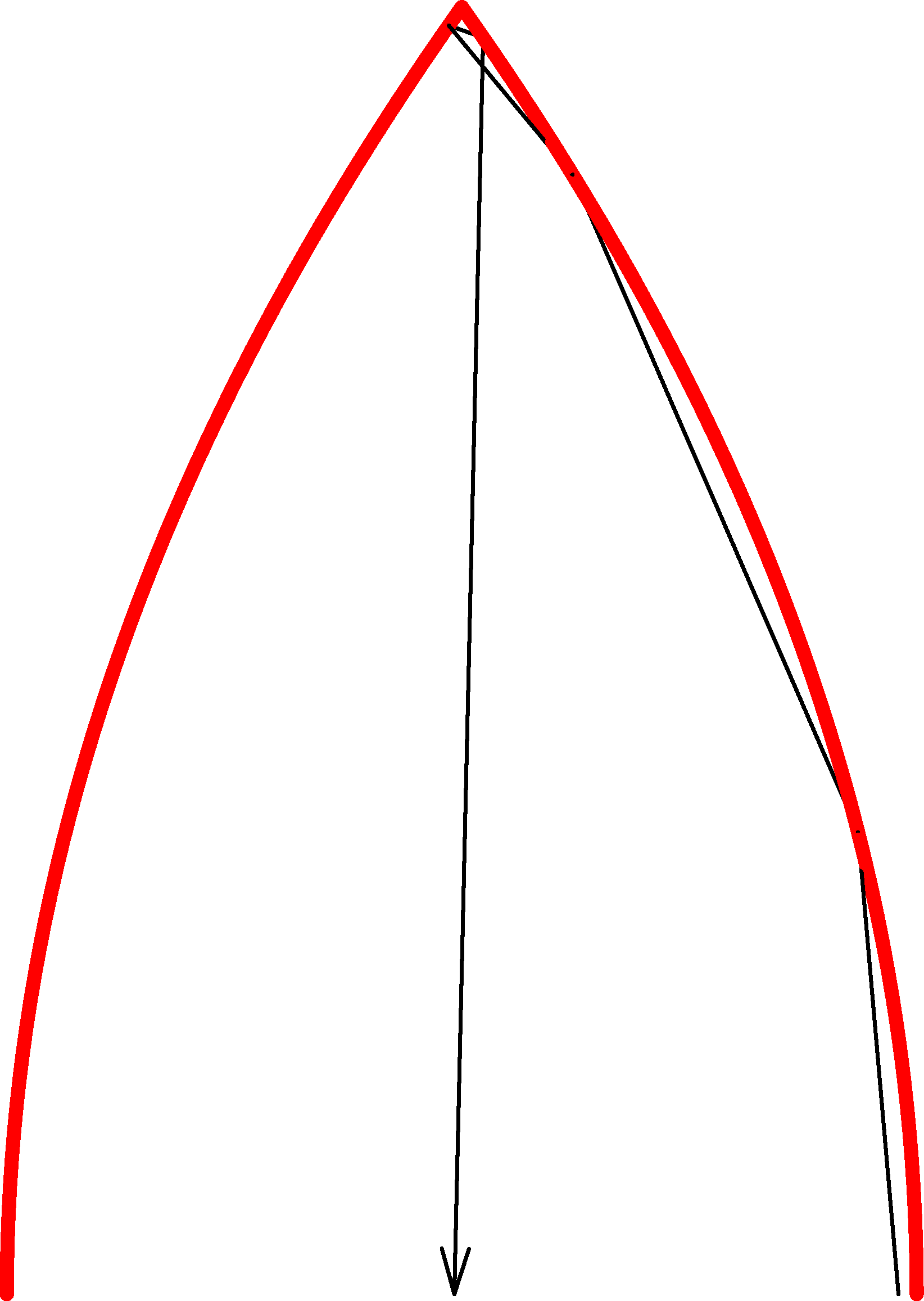} \\
(d) $x=0.3$, $\varphi=75^\circ$.&(e) $x=0.0$, $\varphi=35^\circ$.&(f) $x=0.48$, $\varphi=5^\circ$. \\
\end{tabular}
\caption{Exemplo de trajectórias obtidas com o modelo computacional.}
\label{fig:trajectorias}
\end{center}
\end{figure}

Cada uma das ilustrações da Figura~\ref{fig:trajectorias} reproduz, para a ``Dupla Parábola'', uma trajectória concreta, obtida com o nosso modelo computacional.
É reconfortante verificar que, à excepção da última trajectória, em todas as restantes
a partícula surge à saída da cavidade com uma velocidade quase invertida
relativamente àquela que foi a sua velocidade de entrada. Este é o ``sintoma'' que caracteriza
inequivocamente uma cavidade de óptimo desempenho. Mesmo no caso da trajectória 
da ilustração (f), a direcção da velocidade de saída parece não andar muito distante
da de entrada.

Se analisarmos as cinco primeiras ilustrações, verificamos existir
algo em comum no comportamento da partícula:
para descrever a trajectória, a partícula é sempre sujeita a três reflexões. Esta parece ser uma característica determinante para a aproximação dos ângulos de entrada e de saída. Se, por exemplo, imaginarmos três trajectórias com configurações próximas, respectivamente, das trajectórias (a), (b) e (c), mas com a diferença de não possuírem a terceira reflexão, o resultado será completamente diferente, como facilmente se depreende das ilustrações.
Embora esta convicção seja por enquanto de natureza essencialmente empírica, os resultados
do estudo que se segue vão no sentido de confirmar que uma parte muito significativa das trajectórias ``benignas'' --- aquelas em que os vectores velocidade de entrada e de saída
são quase paralelos; chamemos-lhes assim por representarem contribuições positivas na
maximização da resistência --- subentendem exactamente três reflexões.

\begin{figure}[!hb]
\begin{center}
\includegraphics*[width=0.6\columnwidth]{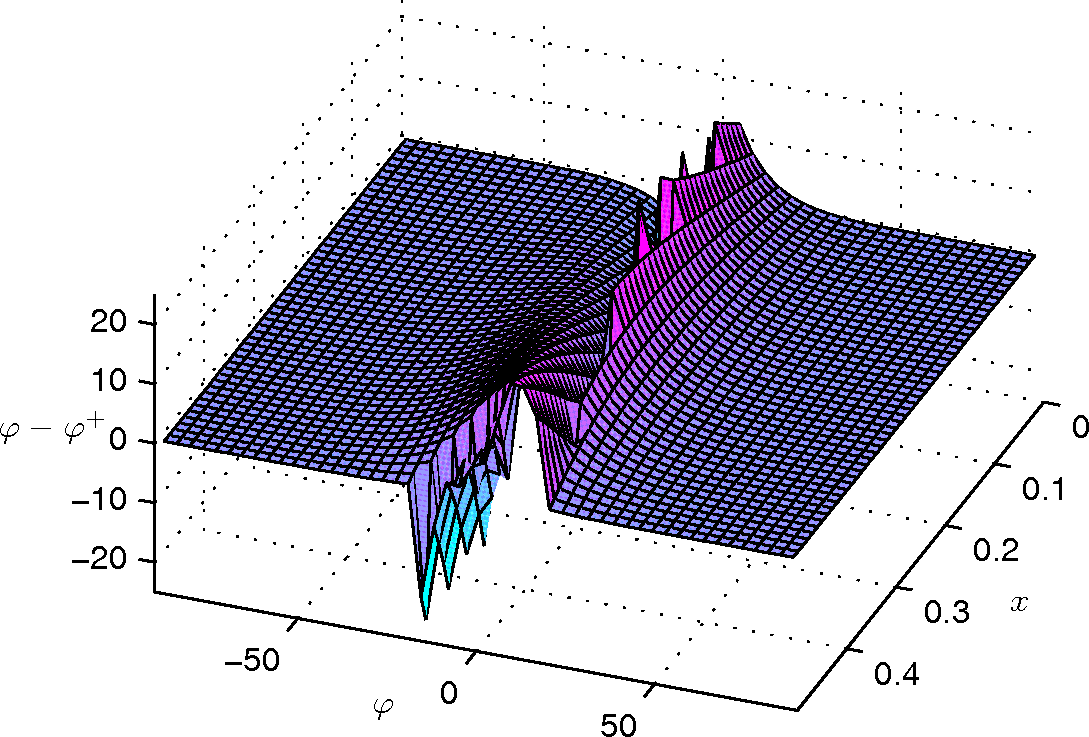}
\caption{Representação gráfica da diferença angular $\varphi-\varphi^+(x,\varphi)$.}
\label{fig:desfasamento}
\end{center}
\end{figure}

Vamos agora tentar interpretar outro tipo de resultados obtidos com o nosso
modelo computacional,
começando pela representação gráfica do valor da diferença entre
os ângulos de entrada e de saída ($\varphi-\varphi^+(x,\varphi))$,
em função da posição de entrada $x$ e do  ângulo de entrada $\varphi$
 --- ver Figura~\ref{fig:desfasamento}.
O gráfico foi produzido usando $100$ valores, uniformemente distribuídos, para cada uma das variáveis $x\in\left(0,0.5\right)$ e $\varphi\in\left(-90^\circ,90^\circ\right)$
--- dada a simetria da cavidade em relação ao eixo $x=0$, obter-se-ia um resultado equivalente
para o intervalo $x\in\left(-0.5,0\right)$.
Este gráfico mostra-nos que, independentemente do valor
da posição de entrada $x$, a diferença angular é aproximadamente nula
para ângulos de entrada $\varphi$ de amplitude elevada,
apresentando valores menos interessantes 
quando $\varphi$ se aproxima de zero.
Portanto, começa-se a perceber que as trajectórias ``benignas'' têm
origem essencialmente em ângulos de entrada de amplitude elevada.

\begin{figure}[!hb]
\begin{center}
\includegraphics*[width=0.6\columnwidth]{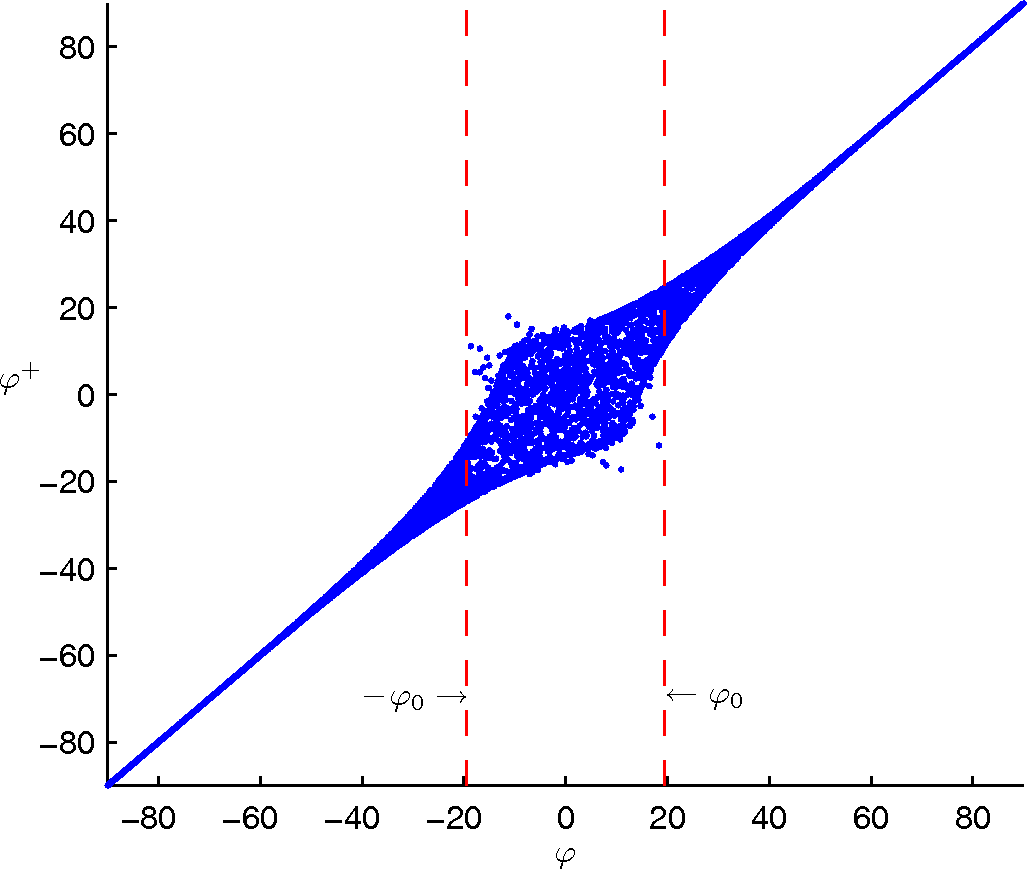}
\caption{Representação no plano cartesiano da distribuição dos pares $(\varphi,\varphi^+)$.}
\label{fig:Dist_Phi_PhiPlus}
\end{center}
\end{figure}

\begin{figure}[!hb]
\begin{center}
\includegraphics*[width=0.6\columnwidth]{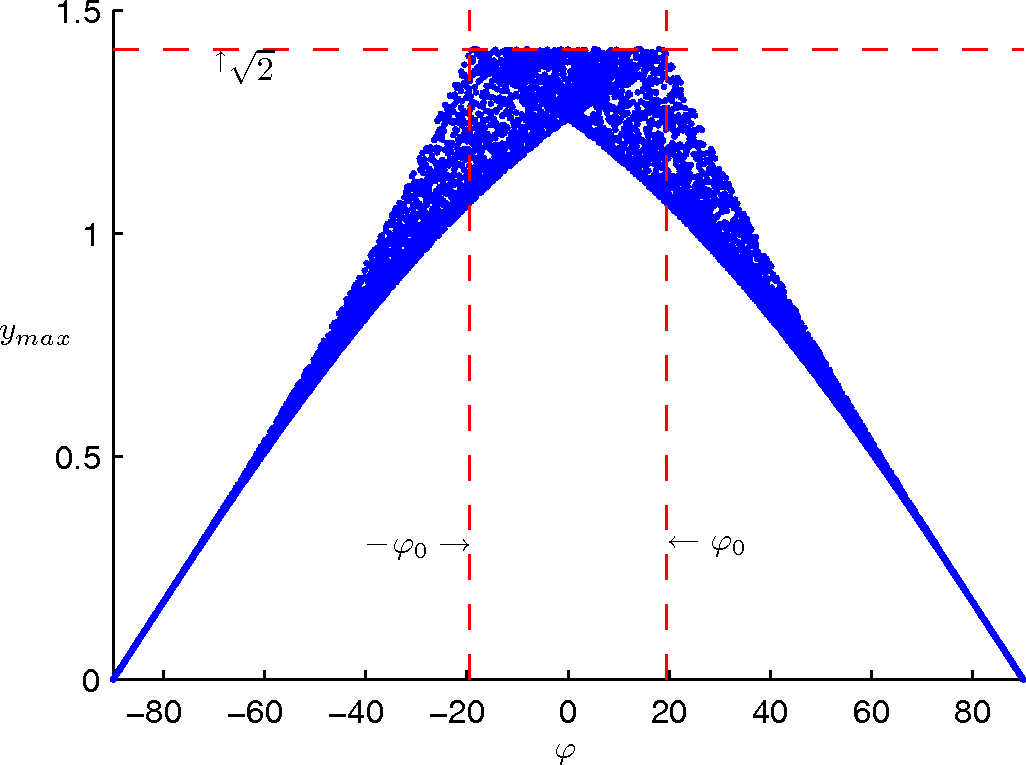}
\caption[Distribuição no plano cartesiano dos pares $(\varphi,y_{max})$.]{Distribuição no plano cartesiano dos pares $(\varphi,y_{max})$,
sendo $y_{max}$ a altura atingida pela partícula.}
\label{fig:Dist_Phi_ymax}
\end{center}
\end{figure}

Os gráficos das figuras~\ref{fig:Dist_Phi_PhiPlus} e \ref{fig:Dist_Phi_ymax},
que se seguem no nosso estudo, foram produzidos
com $10.000$ pares de valores $(x,\varphi)$,
gerados por um processo aleatório de distribuição uniforme.
O primeiro deles mostra a distribuição dos pares $(\varphi,\varphi^+)$ no plano cartesiano.
Os pontos concentram-se nas proximidades da diagonal $\varphi=\varphi^+$, o que é
revelador de um bom comportamento da cavidade.
Também com estes resultados se confirma que a resposta da cavidade se vai deteriorando
quando $\varphi$ se aproxima de zero.

Vejamos o segundo gráfico, expresso na Figura~\ref{fig:Dist_Phi_ymax},
onde se encontra representada no plano cartesiano a distribuição dos pares $(\varphi,y_{max})$,
sendo $y_{max}$ a altura atingida pela partícula.
Pela análise da figura, constatam-se essencialmente duas coisas: que a altura atingida pela partícula quase não depende da posição de entrada $x$, dependendo essencialmente do valor do
ângulo de entrada $\varphi$, em especial para ângulos elevados --- esta relação é
perceptível na Figura~\ref{fig:trajectorias}, quando comparamos a ilustração (a) com a (d) e a (c) com a (e) ---; e que a altura que a partícula pode atingir aumenta com a diminuição da amplitude do ângulo de entrada.

Comparando agora os gráficos \ref{fig:Dist_Phi_PhiPlus} e \ref{fig:Dist_Phi_ymax},
percebe-se que existe uma forte correlação entre o desempenho da cavidade e a altura atingida pela partícula: quanto mais elevada é a altura atingida,
maior é a dispersão dos valores em relação à diagonal $\varphi=\varphi^+$.
Este facto leva-nos a supor que a forma do contorno da cavidade é a ideal junta da sua base, mas
vai deixando de o ser à medida que nos aproximamos do topo da cavidade.

A zona da cavidade que interage com a partícula parece não ser o único factor a condicionar a similitude entre os ângulos $\varphi$ e $\varphi^+$.
Se repararmos no gráfico \ref{fig:Dist_Phi_PhiPlus},
parece existir uma perturbação adicional no comportamento da cavidade
quando a amplitude do ângulo de entrada é inferior a cerca de $20^\circ$,
que leva a que alguns pares $(\varphi,\varphi^+)$ fiquem,
em relação aos restantes, mais dispersos
e mais distantes da diagonal $\varphi^+=\varphi$.
Chamámos já a atenção para a possível importância das 3 reflexões
no grau de aproximação verificado nos ângulos $\varphi$ e $\varphi^+$.
Ocorre-nos por isso a seguinte questão: 
não será precisamente o número de reflexões que, ao se diferenciar das 3
ocorrências, interfere tão negativamente no comportamento da cavidade?
As investigações que se seguem vão demonstrar, entre outras coisas,
que esta nossa suspeita tem fundamento.

A confirmar-se a nossa última conjectura, e atendendo ao aspecto da distribuição dos
pares $(\varphi,\varphi^+)$ no gráfico~\ref{fig:Dist_Phi_PhiPlus},
é expectável que exista para $\varphi$ um intervalo $(-\varphi_0,\varphi_0)\subset (-20^\circ,20^\circ)$
que abarque todas as trajectórias que não sejam formadas por 3 reflexões.
A existência e identificação desse intervalo são garantidas com a demonstração
apresentada na subsecção~\ref{cha:condSuf3col} do postulado que se segue:
\begin{quotation}{\sl
Para ângulos de entrada $\varphi$ superiores (em valor absoluto) a $\varphi_0=\arctan\left(\frac{\sqrt{2}}{4}\right)\simeq 19.47^\circ$,
o número de reflexões a que a partícula é sujeita no interior da cavidade é sempre igual a três,
e ocorrem alternadamente nas faces esquerda e direita da cavidade, para qualquer que 
seja a posição de entrada.
}\end{quotation}

\begin{figure}[!hb]
\begin{center}
\includegraphics*[width=0.6\columnwidth]{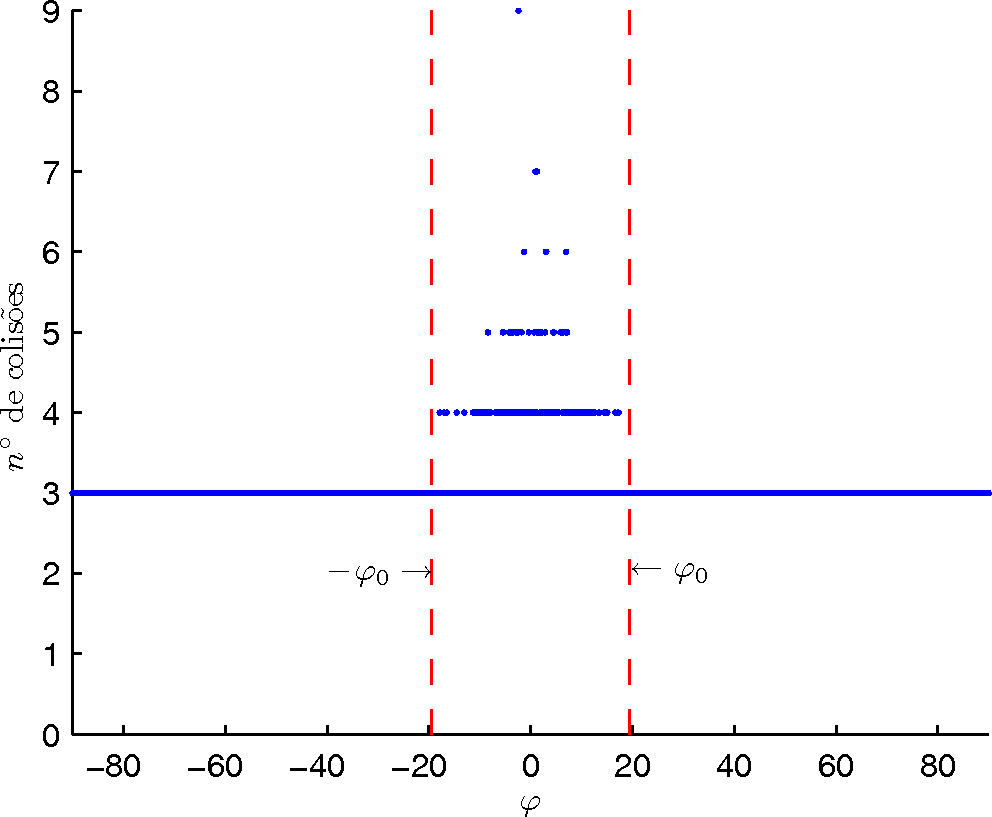}
\caption[Distribuição no plano cartesiano dos pares $(\varphi,nc)$.]{Distribuição no plano cartesiano dos pares $(\varphi,nc)$,
sendo $nc$ o nº de colisões.}
\label{fig:ncolisoes4}
\end{center}
\end{figure}

\begin{figure}[!hb]
\begin{center}
\includegraphics*[width=0.6\columnwidth]{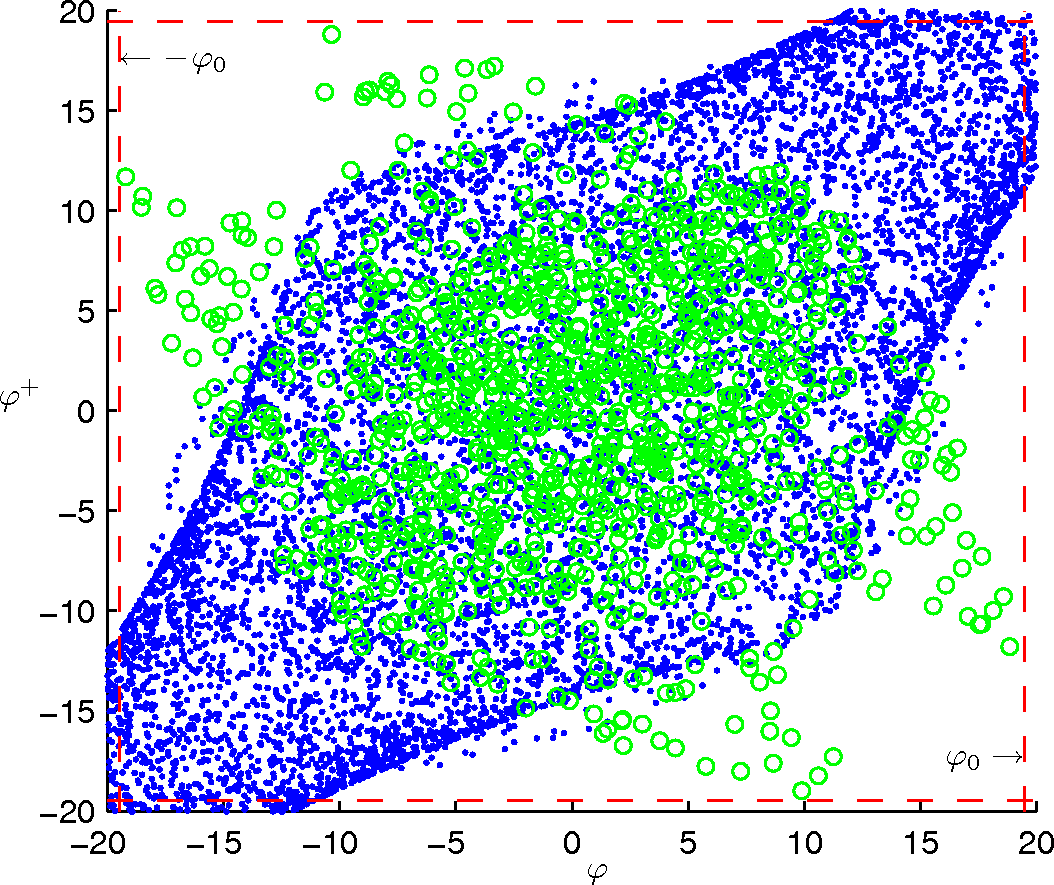}
\caption[Representação dos pares $(\varphi,\varphi^+)$ com 3 e com 4 ou mais reflexões.]{Distribuição dos pares $(\varphi,\varphi^+)$: ($\cdot$) pares
associados a trajectórias com 3 reflexões;
($\circ$) pares associados a trajectórias com 4 ou mais reflexões.}
\label{fig:Dist_Phi_PhiPlus2020}
\end{center}
\end{figure}

De forma a verificarmos que as deduções que fizemos estão efectivamente em concordância com os
resultados numéricos do modelo computacional desenvolvido, apresentamos mais dois
gráficos, Figuras~\ref{fig:ncolisoes4} e \ref{fig:Dist_Phi_PhiPlus2020},
ambos produzidos com $10.000$ pares de valores $(x,\varphi)$,
gerados aleatoriamente com distribuição uniforme.
Como pode ser observado na Figura~\ref{fig:ncolisoes4},
todas as trajectórias com 4 ou mais reflexões (colisões), entre as $10.000$ consideradas,
aconteceram dentro do intervalo $(-\varphi_0,\varphi_0)$.
Fora desse intervalo (para $\left|\varphi\right|>\varphi_0$) as trajectórias são sempre
de três reflexões. Adicionalmente, podemos também verificar não
existir qualquer trajectória com menos de três reflexões.
Provaremos na subsecção~\ref{cha:min3col} ser esta também uma característica
da cavidade.

Não menos elucidativo é o resultado apresentado no gráfico da
Figura~\ref{fig:Dist_Phi_PhiPlus2020},
onde diferenciámos dos restantes os pares $(\varphi,\varphi^+)$ que estão
associados a trajectórias com 3 reflexões.
Constata-se, tal como suspeitávamos, que os pares $(\varphi,\varphi^+)$
que se encontram mais dispersos
e mais distantes da diagonal $\varphi^+=\varphi$, estão
todos eles associados a trajectórias com 4 ou mais reflexões.

Das conclusões a que chegámos podemos de imediato retirar o seguinte corolário:
mesmo nas condições que conduzem ao pior desempenho da cavidade --- 4 ou mais reflexões ---,
a diferença angular $\left|\varphi-\varphi^+\right|$, por maior que seja,
nunca será superior a $2\varphi_0\simeq 38.94^\circ$, valor que é bastante inferior
ao maior ângulo que é possível formar entre dois vectores ($180^\circ$).
A demonstração deste corolário é simples:
como uma trajectória de 4 ou mais reflexões está sempre associada
a um ângulo de entrada $-\varphi_0<\varphi<\varphi_0$, o ângulo de saída situar-se-á
necessariamente no mesmo intervalo;
tendo em conta a propriedade de reversibilidade associada à lei de reflexão que rege
as reflexões, se por absurdo admitíssemos $\left|\varphi^+\right|>\varphi_0$, ao
invertermos o sentido de marcha da partícula, passaríamos a ter uma trajectória
de mais de 3 reflexões com um ângulo de entrada $\varphi^+$ situado fora do
intervalo $(-\varphi_0,\varphi_0)$, o que entraria em contradição
com o postulado inicial. 
Os resultados apresentados na Figura~\ref{fig:Dist_Phi_PhiPlus2020}
estão em concordância com este novo facto,
uma vez que todos os pares $(\varphi,\varphi^+)$ com mais de três reflexões
parecem situar-se no interior da região $(-\varphi_0,\varphi_0)\times(-\varphi_0,\varphi_0)$.

Resumindo:
\begin{itemize}
\item A altura que a partícula atinge no interior da cavidade interfere
negativamente na aproximação dos ângulos $\varphi$ e $\varphi^+$, levando-nos
a pensar que a forma da cavidade é a ideal, ou muito próxima da ideal, junto à sua base, mas
vai perdendo qualidades em zonas mais elevadas da cavidade;
\item O maior desfazamento entre $\varphi$ e $\varphi^+$ ocorre em trajectórias de 4 ou mais reflexões;
\item Verifica-se, em todo o intervalo de variação de $\varphi$, uma grande predominância das trajectórias com 3 reflexões;
\item Não existem trajectórias com menos de 3 reflexões;
\item O ângulo crítico $\varphi_0$ tem o valor: $\varphi_0=\arctan\left(\frac{\sqrt{2}}{4}\right)\simeq 19.47^\circ$;
\item Fora do intervalo $(-\varphi_0,\varphi_0)$, todas as trajectórias são de 3 reflexões;
\item Em trajectórias com 4 ou mais reflexões, a diferença angular
é delimitada por $2\varphi_0$: $\left|\varphi-\varphi^+\right|<2\varphi_0$.
\end{itemize}


\subsection{Uma condição suficiente para a ocorrência de três reflexões}
\label{cha:condSuf3col}


\begin{theorem}
\label{teor:3ref}
Para ângulos de entrada $\varphi$ superiores (em valor absoluto) a $\varphi_0=\arctan\left(\frac{\sqrt{2}}{4}\right)\simeq 19.47^\circ$, o número de reflexões a que a partícula é sujeita no interior da cavidade Dupla Parábola é sempre igual a três, e ocorrem alternadamente nas faces esquerda e direita da cavidade, para qualquer que seja a posição de entrada.
\end{theorem}

\begin{figure}[!hb]
\begin{center}
\includegraphics*[width=0.8\columnwidth]{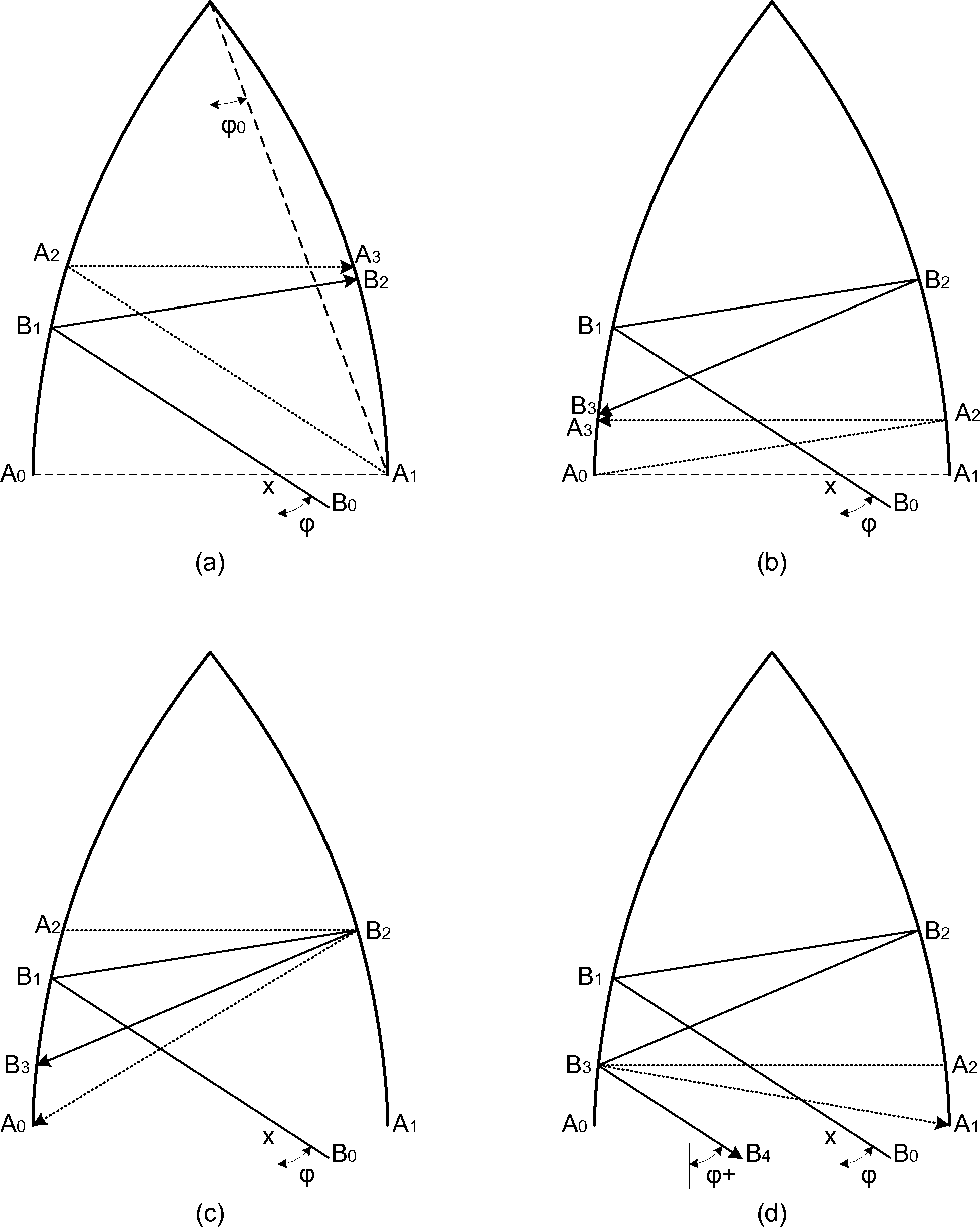}
\caption[Ilustrações para estudo das trajectórias com
ângulos de entrada $\varphi>\varphi_0$.]{Conjunto de ilustrações para estudo da trajectória de partículas com
ângulos de entrada $\varphi>\varphi_0 \simeq 19.47^\circ$,
na cavidade ``Dupla Parábola''.}
\label{fig:parabOpt3col}
\end{center}
\end{figure}

\begin{figure}[!hb]
\begin{center}
\includegraphics*[width=0.8\columnwidth]{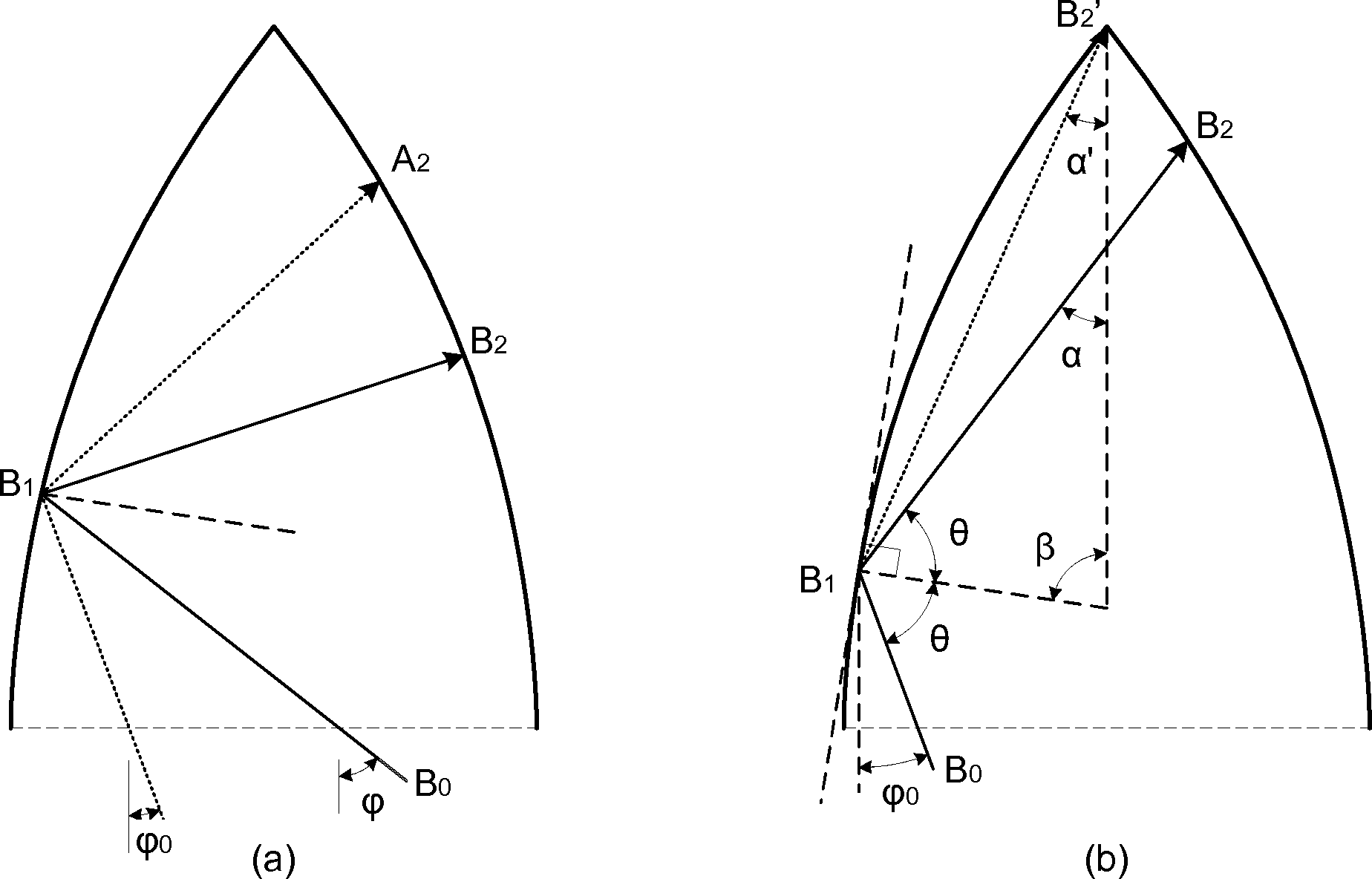}
\caption{Ilustrações para estudo da segunda reflexão.}
\label{fig:parabOpt3colA}
\end{center}
\end{figure}

\begin{figure}[!hb]
\begin{center}
\includegraphics*[width=0.8\columnwidth]{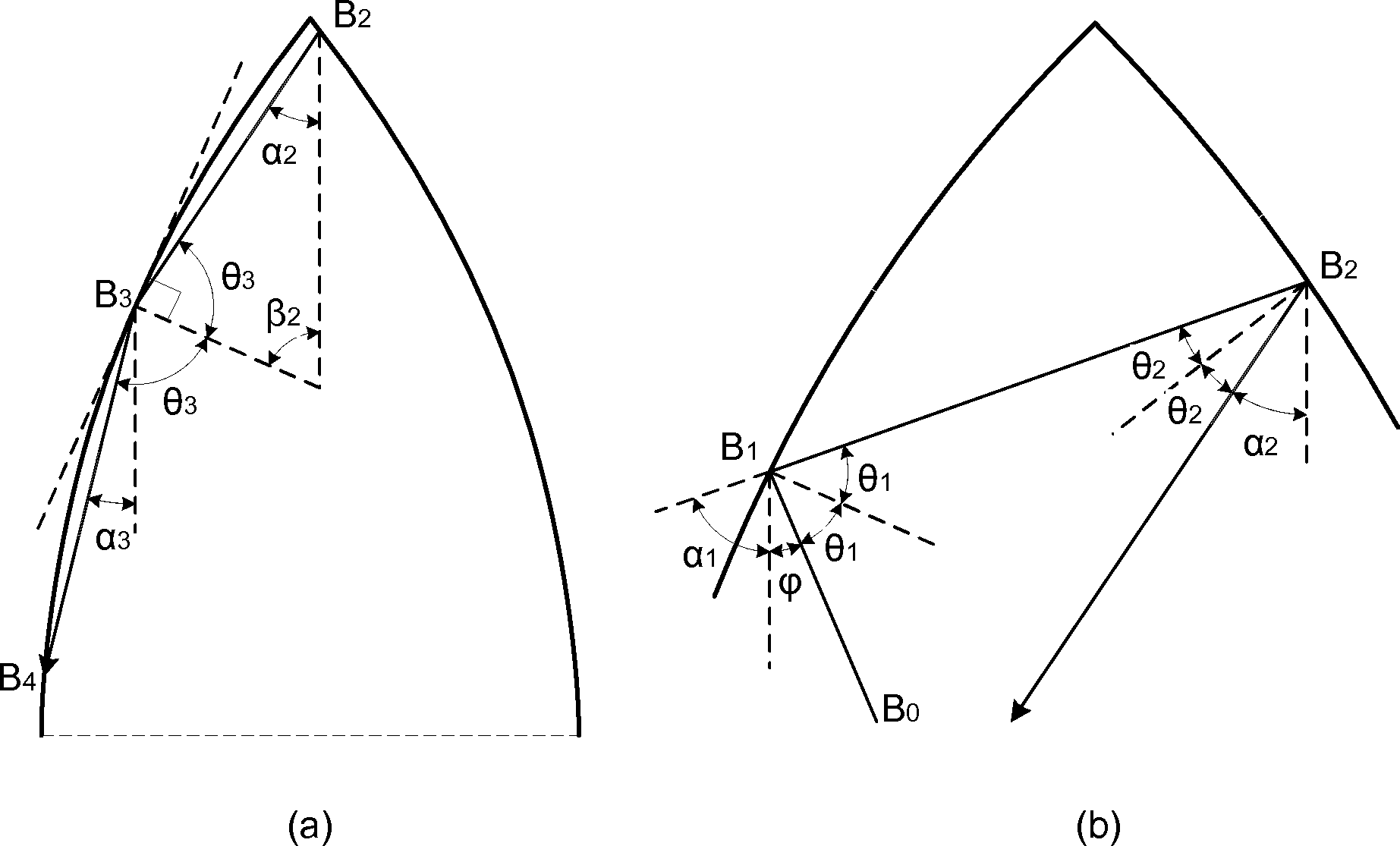}
\caption{Ilustrações para estudo de uma hipotética quarta reflexão.}
\label{fig:parabOpt3colB}
\end{center}
\end{figure}

Para demonstrarmos o teorema que acabámos de enunciar, estudaremos a trajectória de uma partícula passo a passo, desde o momento em que entra na cavidade até ao momento em que a abandona,
e usaremos as ilustrações das figuras~\ref{fig:parabOpt3col}--\ref{fig:parabOpt3colB} para nos auxiliarem nessa demonstração.

Considere-se uma partícula que entra na cavidade em $(x,0)$,
com o vector velocidade a formar um ângulo $\varphi$ com o eixo vertical, tal como se encontra representado nas ilustrações da Figura~\ref{fig:parabOpt3col}, onde assumimos que o eixo de simetria da cavidade é o eixo dos $y$ e que a sua base
$\overline{A_0A_1}$ assenta no eixo dos $x$. Assim, a posição da partícula à entrada da cavidade assume apenas valores no intervalo $\left(-\frac{1}{2},\,\frac{1}{2}\right)\times\{0\}$.  

Dada a simetria da cavidade em relação ao seu eixo vertical, será suficiente analisar
o seu comportamento para $\varphi_0 <\varphi< 90^\circ$. As conclusões a que chegarmos
serão assim igualmente válidas para $-90^\circ<\varphi< -\varphi_0$.

Analisemos então em detalhe e separadamente cada um dos subtrajectos que compõem toda a trajectória descrita pelo movimento da partícula no interior da cavidade.

\vspace{0.5cm}
\noindent
\emph{Subtrajecto $\overrightarrow{B_0B_1}$}

Para $\varphi>\varphi_0$, temos a garantia de que a primeira reflexão ocorre na curva parabólica do lado esquerdo da cavidade, tal como pode ser facilmente deduzido a partir da ilustração (a).
Para que a partícula colida na curva esquerda bastaria que o ângulo $\varphi$ fosse superior a $\arctan\left(\frac{x \sqrt{2}}{2}\right)$, grandeza que tem como majorante
$\varphi_0=\arctan\left(\frac{\sqrt{2}}{4}\right)$. Teremos assim o trajecto inicial da cavidade
representado na ilustração (a) pelo vector $\overrightarrow{B_0B_1}$.

\vspace{0.5cm}
\noindent
\emph{Subtrajecto $\overrightarrow{B_1B_2}$}
\label{pg:subtrajB1B2}

Após colidir em $B_1$, de acordo com a lei de reflexão, a partícula segue pelo
trajecto $\overrightarrow{B_1B_2}$.
Demonstremos que $\overrightarrow{B_1B_2}$ tem o
sentido ascendente --- ilustração (a).
Tracemos o segmento de recta $\overline{A_1A_2}$, paralelo ao trajecto inicial da partícula $\overline{B_0B_1}$
e que passe pelo foco da parábola esquerda ($A_1$).
Pela propriedade focal dessa parábola, uma partícula que tome o subtrajecto $\overrightarrow{A_1A_2}$, após a reflexão em $A_2$, seguirá numa direcção horizontal $\overline{A_2A_3}$ (prosseguindo depois o seu trajecto, após nova reflexão, em direcção ao foco $A_0$ da segunda parábola).
Ocorrendo a primeira reflexão da partícula em $B_1$, um ponto da curva necessariamente posicionado abaixo de $A_2$, o trajecto $\overrightarrow{B_1B_2}$, que seguirá
de imediato, será no sentido ascendente, pois a derivada $\frac{\dd y}{\dd x}$ da curva nesse ponto ($B_1$) é superior à derivada em $A_2$,
onde a trajectória que se seguia era horizontal.

Embora saibamos já que $\overrightarrow{B_1B_2}$ tem o sentido ascendente,
ainda nada nos garante que a segunda reflexão aconteça
necessariamente na parábola do lado direito. 
Se conseguirmos verificar que para $\varphi=\varphi_0$ a segunda reflexão
é sempre no lado direito,
para qualquer que seja a posição de entrada $x$, então,
por maioria de razão, o mesmo sucederá para
qualquer valor $\varphi>\varphi_0$.
Esta premissa pode ser facilmente aceite com o auxílio da ilustração (a) da Figura~\ref{fig:parabOpt3colA}: para qualquer valor de $\varphi>\varphi_0$, com
a primeira reflexão num dado ponto $B_1$, é
sempre possível traçarmos uma trajectória para $\varphi=\varphi_0$ que apresente
a primeira reflexão no mesmo ponto $B_1$; sendo a segunda reflexão
na curva do lado direito para o caso $\varphi=\varphi_0$, necessariamente o mesmo
acontecerá para a trajectória com $\varphi>\varphi_0$, pois o ângulo de
reflexão será menor neste segundo caso, tal como
se ilustra na figura.
Por conseguinte, bastar-nos-há provar para $\varphi=\varphi_0$, que a 
segunda reflexão ocorre sempre na parábola do lado direito,
para que o mesmo fique provado para qualquer
que seja o $\varphi>\varphi_0$.

Na ilustração (b) da Figura~\ref{fig:parabOpt3colA}
encontra-se representada a trajectória até à segunda
reflexão de uma partícula com ângulo de entrada $\varphi_0$
($\overrightarrow{B_0B_1}$ e $\overrightarrow{B_1B_2}$).
Como se depreende da ilustração, a reflexão $B_2$ só
acontecerá na curva do lado esquerdo se o ângulo
$\alpha$ for menor que $\alpha'$. 
Determinemos o valor dos dois ângulos.

Sendo $(x_1,y_1)$ as coordenadas do ponto $B_1$,
teremos $\tan(\alpha')=\frac{-x_1}{\sqrt{2}-y_1}$,
logo
\begin{equation}
\label{eq:alphalinha}
\alpha'
=\arctan\left(\frac{-(y_1^2/4-1/2)}{\sqrt{2}-y_1}\right)
=\arctan\left(\frac{(2-y_1^2)/4}{\sqrt{2}-y_1}\right)
=\arctan\left(\frac{\sqrt{2}+y_1}{4}\right)\text{.}
\end{equation}
Para chegarmos ao valor de $\alpha$ resolvemos
o sistema de três equações, de incógnitas $\alpha$, $\theta$ e $\beta$,
que se retira directamente da geometria da própria figura
\begin{equation*}
\left\{
\begin{array}{l}
\alpha + \beta + \theta = \pi\\
\beta = \varphi_0+\theta\\
\arctan \left(\frac{1}{2}y_1\right)+\varphi_0+\theta = \frac{\pi}{2}
\end{array}
\right.
\end{equation*}

A linha tangente à curva em $B_1$ faz com a vertical um ângulo
cuja tangente tem por valor a derivada 
$\frac{\dd x}{\dd y}$ da curva nesse ponto (em $y=y_1$), onde $\frac{\dd x}{\dd y}
=\frac{\dd}{\dd y}(\frac{1}{4}y^2-\frac{1}{2})=\frac{1}{2}y$. Por
isso, esse ângulo surge representado na terceira das equações pela
grandeza $\arctan \left(\frac{1}{2}y_1\right)$.

Resolvendo o sistema, obtém-se para $\alpha$ o seguinte resultado
\begin{equation}
\label{eq:alpha}
\alpha = \varphi_0+2\arctan \left(\frac{y_1}{2}\right)
=\arctan\left(\frac{\sqrt{2}}{4}\right)+2\arctan \left(\frac{y_1}{2}\right)
\text{.}
\end{equation}

Provemos finalmente que $\alpha>\alpha'$, para qualquer que seja $y_1\in \, \left(0,\sqrt{2}\right)$.
Das equações~\eqref{eq:alpha} e \eqref{eq:alphalinha},
será equivalente a provarmos
\begin{eqnarray*}
\arctan\left(\frac{\sqrt{2}}{4}\right)+2\arctan \left(\frac{y_1}{2}\right)>
\arctan \left(\frac{\sqrt{2}+y_1}{4}\right)\text{.}
\end{eqnarray*}
Dado que $0<y_1<\sqrt{2}$, ambos os membros da inequação representam
ângulos situados no primeiro quadrante do círculo trigonométrico.
Por isso podemos manter a inequação para a tangente dos respectivos ângulos.
Aplicando a tangente a ambos os membros, depois de efectuadas
algumas simplificações trigonométricas, chegamos á seguinte relação
\begin{equation*}
\frac{1}{4}\,\frac{4\sqrt{2}-\sqrt{2}y_1^2+16 y_1}{4-y_1^2-\sqrt{2} y_1}>
\frac{\sqrt{2}+y_1}{4} \text{,} 
\end{equation*}
que, com simplificações algébricas adicionais, toma a forma
\begin{equation*}
\frac{y_1\left(\sqrt{2}y_1+14+y_1^2\right)}{4-y_1^2-\sqrt{2} y_1}>0 \text{.}
\end{equation*}
Como $0<y_1<\sqrt{2}$, facilmente se constata que tanto o numerador
como o denominador da fracção presente nesta última inequação
são grandezas positivas. Logo $\alpha>\alpha'$,
que contraria a condição que era necessária para que a reflexão
$B_2$ ocorresse na curva do lado esquerdo,
ficando assim provado,
como pretendíamos, que em nenhuma situação a reflexão $B_2$
da ilustração (b) da Figura~\ref{fig:parabOpt3colA} acontece
na curva do lado esquerdo. Por maioria de razão, podemos
então também
concluir que o mesmo sucede para qualquer que seja $\varphi>\varphi_0$:
a segunda reflexão da partícula ocorre sempre na parábola do lado direito. 

\vspace{0.5cm}
\noindent
\emph{Subtrajecto $\overrightarrow{B_2B_3}$}
\label{pg:subtrajB2B3}

Demonstremos que o subtrajecto $\overrightarrow{B_2B_3}$
tem o sentido descendente --- ilustração (b) da Figura~\ref{fig:parabOpt3col}.
Imagine-se, para o efeito, um subtrajecto $\overrightarrow{A_0A_2}$, paralelo a $\overline{B_1B_2}$ e que passe no foco $A_0$. O subtrajecto $\overrightarrow{A_2A_3}$ que se
seguiria à reflexão em $A_2$ --- um ponto da parábola do lado direito situado abaixo de $B_2$ --- seria horizontal. Sendo a derivada da curva em $A_2$ superior ao valor da derivada em $B_2$,
o subtrajecto $\overline{B_2B_3}$ será necessariamente de natureza descendente.

Ainda que já saibamos que o subtrajecto é descendente, ainda não mostrámos 
que esse subtrajecto em nenhuma situação conduz a partícula directamente
para a saída da cavidade. Segue-se então a demonstração de que a reflexão $B_3$
ocorre sempre numa posição superior a $A_0$ --- ilustração (c) da Figura~\ref{fig:parabOpt3col}.
Tracemos $\overline{A_2B_2}$, um segmento de recta horizontal que passe
no ponto de reflexão $B_2$. Se a partícula seguisse esse trajecto, colidiria
no mesmo ponto $B_2$, mas dirigir-se-ia para $A_0$. Logo, pela lei de reflexão,
$B_3$ terá que estar acima de $A_0$, pois $\overline{B_1B_2}$ faz um ângulo com o vector normal à curva em $B_2$ menor que o formado pelo segmento $\overline{A_2B_2}$.

\vspace{0.5cm}
\noindent
\emph{Subtrajecto $\overrightarrow{B_3B_4}$}

Vamos agora mostrar que o subtrajecto que se segue à reflexão em $B_3$ cruza o segmento $\overline{A_0A_1}$, isto é, direcciona-se para fora da cavidade --- ilustração (d) da Figura~\ref{fig:parabOpt3col}.
Tracemos então $\overline{A_2B_3}$, um segmento de recta horizontal que passe
no ponto de reflexão $B_3$. Se a partícula seguisse esse trajecto, colidia
em $B_3$ e dirigir-se-ia para $A_1$. Logo, pela lei de reflexão,
a recta onde assenta o subtrajecto $\overrightarrow{B_3B_4}$ terá necessariamente
que passar abaixo de $A_1$,
pois $\overline{B_2B_3}$ faz um ângulo com o vector normal à curva em $B_3$ maior que o formado pelo segmento $\overline{A_2B_3}$.
Mostrámos que o subtrajecto cruza o eixo dos $x$ num ponto situado à esquerda de $A_1$, mas
ainda não mostrámos que ocorre à direita de $A_0$. Para tal, teremos que demonstrar que a terceira é a última das reflexões, isto é, que em nenhuma situação ocorre uma quarta reflexão na parábola do lado esquerdo. Segue-se essa demonstração, de todas a mais complexa e a mais demorada.

Para provarmos que a seguir à terceira reflexão não ocorre qualquer outra colisão na parábola
esquerda, vamos mostrar que uma quarta colisão --- representada por $B_4$ na ilustração (a) da Figura~\ref{fig:parabOpt3colB} --- tem sempre origem num ângulo de entrada $\varphi$ inferior a $\varphi_0$.
Iremos assim estudar o trajecto da partícula na ordem inversa à sua progressão: começamos por admitir a existência do subtrajecto $\overrightarrow{B_3B_4}$ da ilustração (a) e analisaremos as suas implicações em todo o trajecto precedente.
 
Na ilustração (a) da Figura~\ref{fig:parabOpt3colB} encontram-se representados os
subtrajectos $\overrightarrow{B_2B_3}$ e $\overrightarrow{B_3B_4}$.
Comecemos por relacionar
$\alpha_2$ com $\alpha_3$, os ângulos que os vectores $\overrightarrow{B_2B_3}$ e $\overrightarrow{B_3B_4}$, respectivamente, formam com o eixo vertical.
Para o efeito resolvemos
o sistema de três equações, de incógnitas $\alpha_2$, $\theta_3$ e $\beta_2$,
que se retira da geometria da figura,\footnote{As variáveis denotadas por $x_i$ e $y_i$,
com $i=1,\ldots,4$, representam as coordenadas do $i$-ésimo ponto de reflexão, identificado por $B_i$.}
\begin{equation*}
\left\{
\begin{array}{l}
\theta_3 =\beta_2 + \alpha_3\\
\arctan \left(\frac{1}{2}y_3\right)+\beta_2 = \frac{\pi}{2}\\
\alpha_2 +\theta_3+\beta_2 = \pi
\end{array}
\right.
\end{equation*}
obtendo-se
\begin{equation*}
\label{eq:alpha2}
\alpha_2 =2\arctan \left(\frac{y_3}{2}\right)-\alpha_3
\text{,}
\end{equation*}
em que $\arctan\left(\frac{1}{2}y_3\right)$ é o ângulo que a recta tangente
à curva em $B_3$ faz com a vertical --- o declive da recta tangente é dado por
$\frac{\dd x}{\dd y}\left.\right|_{y=y_3}
=\frac{\dd}{\dd y}(\frac{1}{4}y^2-\frac{1}{2})\left.\right|_{y=y_3}
=\frac{1}{2}y_3$.
Por sua vez, o ângulo $\alpha_3$ pode ser expresso da seguinte forma
\begin{equation*}
\label{eq:alpha3}
\alpha_3 =\arctan \left(\frac{x_3-x_4}{y_3-y_4}\right)
=\arctan \left(\frac{\frac{1}{4}y_3^2-\frac{1}{4}y_4^2}{y_3-y_4}\right)
=\arctan \left(\frac{y_3+y_4}{4}\right)
\text{,}
\end{equation*}
o que nos permite escrever $\alpha_2$ em função unicamente das
ordenadas $y_3$ e $y_4$ dos extremos do vector $\overrightarrow{B_3B_4}$,
\begin{equation}
\label{eq:alpha2b}
\alpha_2 =2\arctan \left(\frac{y_3}{2}\right)-\arctan \left(\frac{y_3+y_4}{4}\right)
\text{.}
\end{equation}

Para conseguirmos provar o que pretendemos --- impossibilidade de ocorrência da reflexão $B_4$ --- precisamos de encontrar um minorante para a ordenada da posição onde ocorre cada uma das quatro reflexões, ou seja, determinar $\left\{y_1^*,y_2^*,y_3^*,y_4^*\right\}$, tal que 

\begin{equation}
\label{eq:minorantes}
y_1\geq y_1^*,\;y_2\geq y_2^*,\;y_3\geq y_3^*,\;y_4\geq y_4^*,\;
\;\forall (\varphi,x)\in\left(\varphi_0,\frac{\pi}{2}\right)\times\left(-\frac{1}{2},\frac{1}{2}\right)\text{.}
\end{equation}

Facilmente se percebe que $y_4^*=0$. Vamos então determinar os outros
três minorantes, começando por $y_2^*$.

Sabemos que $0<y_4<y_3$; logo, de \eqref{eq:alpha2b} retiramos que
\begin{equation*}
\label{eq:alpha2Limites}
\arctan \left(\frac{y_3}{2}\right)<\alpha_2 < 2\arctan \left(\frac{y_3}{2}\right)-\arctan \left(\frac{y_3}{4}\right)
\text{.}
\end{equation*}
Atendendo a que $\alpha_2$ se situa no primeiro quadrante do círculo trigonométrico,
podemos manter as desigualdades para a tangente dos respectivos ângulos. Após algumas
simplificações algébricas, obtemos
\begin{equation}
\label{eq:tanAlpha2Limites}
\frac{y_3}{2}<\tan(\alpha_2) < \frac{y_3\left(12+y_3^2\right)}{16}
\text{.}
\end{equation}

A equação da recta que liga $B_2$ a $B_3$ toma a forma
\begin{equation*}
x=m(y-y_3)+x_3,
\end{equation*}
com $m=\tan(\alpha_2)$ e $x_3=\frac{1}{4}y_3^2-\frac{1}{2}$. Como estamos interessados
em encontrar a ordenada do ponto de intercepção dessa recta com a curva parabólica situada no lado direito, de equação
\begin{equation*}
x=-\frac{1}{4}y^2+\frac{1}{2}
\text{,}
\end{equation*}
temos de resolver a equação de segundo grau, na variável $y$, que resulta
da eliminação da variável $x$ por combinação das duas equações anteriores.
A ordenada $y_2$, da segunda reflexão, sendo a raíz positiva da equação,
toma a forma
\begin{equation*}
y_2 = -2 m+\sqrt{4 m^2+4 m y_3 - y_3^2+4}
\text{ .}
\end{equation*}
A grandeza $y_2$ é expressa em função de duas variáveis, $m$ e $y_3$,
que como sabemos assumem apenas valores positivos.
De forma a aceitarmos mais facilmente as deduções que iremos fazer no encalço de $y_2^*$,
imaginemos, sem qualquer perda de generalidade, que $y_3$ é um valor fixo.
Comecemos por mostrar que a derivada de $y_2$ em ordem à variável $m$,
\begin{equation}
\label{eq:derivaday2}
\frac{\dd y_2}{\dd m}=
\frac{2\left(2m+y_3 -\sqrt{4 m^2+4m y_3 -y_3^2 + 4}\right)}{\sqrt{4 m^2+4m y_3 -y_3^2 + 4}}
\text{ ,}
\end{equation}
tem um valor negativo para qualquer que seja o valor de $y_3$.
Como $y_3<\sqrt{2}$, forçosamente $y_3^2<4$,
logo os dois radicandos $(4m^2+4my_3-y_3^2+4)$ presentes na equação~\eqref{eq:derivaday2} têm sempre um valor positivo.
A restrição $y_3<\sqrt{2}$ permite-nos ainda deduzir sucessivamente
as seguintes desigualdades
\begin{eqnarray*}
y_3^2<2 \Leftrightarrow 2 y_3^2<4 \Leftrightarrow y_3^2<4-y_3^2
\Leftrightarrow 4 m^2+4 m y_3 + y_3^2<4 m^2+4 m y_3 +4-y_3^2\\
\Leftrightarrow  \left(2 m+ y_3\right)^2<4 m^2+4 m y_3 -y_3^2+4 
\Leftrightarrow 2 m+ y_3<\sqrt{4 m^2+4 m y_3 -y_3^2+4} 
\text{ .}
\end{eqnarray*}
Esta última desigualdade confirma que $\frac{\dd y_2}{\dd m}<0$, para qualquer
que seja o $y_3$. Assim, o valor $y_2$ é tanto menor quanto maior for o valor de $m$.
Como se depreende de \eqref{eq:tanAlpha2Limites}, $m<M=\frac{y_3\left(12+y_3^2\right)}{16}$, logo
\begin{equation*}
y_2 > -2 M+\sqrt{4 M^2+4 M y_3 - y_3^2+4}
\text{ .}
\end{equation*}
Substituindo $M$, obtém-se, após algumas simplificações,
\begin{equation}
\label{eq:y2minor}
y_2 > f(y_3), \text{ com } f(y_3)=-\frac{3}{2} y_3-\frac{1}{8} y_3^3+\frac{1}{8} \sqrt{272 y_3^2+40 y_3^4+y_3^6+256}
\text{ .}
\end{equation}
Para encontrarmos o valor mínimo de $f(y_3)$ começamos por derivar,
\begin{equation*}
\frac{\dd}{\dd y_3}f(y_3)= \frac{272 y_3+80 y_3^3+3 y_3^5-(12+3 y_3^2) \sqrt{272 y_3^2+40 y_3^4+y_3^6+256}}
{8 \sqrt{272 y_3^2+40 y_3^4+y_3^6+256}}
\text{.}
\end{equation*}
Sendo os radicandos claramente positivos, apenas temos que nos preocupar com o numerador da fracção. Encontrarmos as raízes da função derivada $\frac{\dd}{\dd y_3}f(y_3)$ equivale
por isso a resolvermos a equação
\begin{equation*}
\left(272 y_3+80 y_3^3+3 y_3^5\right)^2=\left(12+3 y_3^2\right)^2 \left(272 y_3^2+40 y_3^4+y_3^6+256\right)
\text{,}
\end{equation*}
que pode ser simplificada na seguinte:
\begin{equation*}
2304-1024 y_3^2-992 y_3^4-160 y_3^6-3 y_3^8=0 \text{.}
\end{equation*}
Esta equação polinomial tem uma única raíz real positiva, de valor
\begin{equation*}
\tilde{y}_3=\frac{2}{3} \sqrt{-51 + 6 \sqrt{79}}\simeq 1.017 \text{,}
\end{equation*}
significando que $f(y_3)$ tem um mínimo global em $\tilde{y}_3$, pois, como
mostramos a seguir, $\frac{\dd^2}{\dd y_3^2}f(y_3)>0$ e a função
não apresenta outros pontos de estacionaridade.

Mostremos então que $\frac{\dd^2}{\dd y_3^2}f(y_3)>0$, com
\begin{eqnarray}
\label{eq:y2segDeriv}
\frac{\dd^2}{\dd y_3^2}f(y_3)= \frac{
18240 y_3^4+2960 y_3^6+180 y_3^8+3 y_3^{10}+34816+30720 y_3^2
}{
4 (272 y_3^2+40 y_3^4+y_3^6+256)^\frac{3}{2}
} \nonumber\\
-\frac{
(816 y_3^3+120 y_3^5+3 y_3^7+768 y_3) \sqrt{272 y_3^2+40 y_3^4+y_3^6+256}
}{
4 (272 y_3^2+40 y_3^4+y_3^6+256)^\frac{3}{2}
}
\text{.}
\end{eqnarray}
Mostrarmos que $\frac{\dd^2}{\dd y_3^2}f(y_3)>0$ equivale a mostrarmos que
o numerador da primeira fracção é superior ao numerador da segunda,
em \eqref{eq:y2segDeriv}. Depois de elevarmos os dois termos ao quadrado, chegamos
à inequação
\begin{eqnarray*}
-3 y_3^{16}+36 y_3^{14}+9464 y_3^{12}+191616 y_3^{10}+1514112 y_3^8+5817344 y_3^6
\nonumber\\
+13535232 y_3^4+15532032 y_3^2+9469952>0 
\text{.}
\end{eqnarray*}
Facilmente comprovamos a veracidade desta relação, dado que temos um único
termo negativo ($-3 y_3^{16}$) que, por exemplo, é inferior em valor absoluto
ao termo constante ($9469952$),
\begin{equation*}
y_3<\sqrt{2} \Rightarrow 3 y_3^{16}<768<9469952 \text{.}
\end{equation*}
Fica assim completa a demonstração de que $f(y_3)$ tem um mínimo global
em $\tilde{y}_3$, de valor
\begin{equation*}
f(\tilde{y}_3)=\frac{8}{9} \sqrt{-51 + 6 \sqrt{79}} \text{.}
\end{equation*}
Portanto, de \eqref{eq:y2minor}, concluímos finalmente que
\begin{equation}
\label{eq:y2min}
y_2>y_2^*=\frac{8}{9} \sqrt{-51 + 6 \sqrt{79}} \simeq 1.356 \text{.}
\end{equation}
Está então encontrado um minorante para a altura da segunda reflexão $B_2$
(ilustração (a) da Figura~\ref{fig:parabOpt3colB}). Determinemos agora 
$y_3^*$, um minorante para a altura da terceira reflexão $B_3$.

Para que a reflexão $B_2$ ocorra na parábola do lado direito é necessário
que o ângulo $\alpha_2$ seja maior que o ângulo formado entre o eixo vertical
e o segmento de recta que une $B_3$ com o vértice superior da cavidade,
\begin{equation}
\label{eq:alpha2Maj}
\alpha_2>
\arctan \left(\frac{0-x_3}{\sqrt{2}-y_3}\right)
=\arctan \left(\frac{-\frac{1}{4}y_3^2+\frac{1}{2}}{\sqrt{2}-y_3}\right)
=\arctan \left(\frac{\sqrt{2}+y_3}{4}\right)
\text{.}
\end{equation}
Esta inequação, conjuntamente com a segunda relação de desigualdade de
\eqref{eq:tanAlpha2Limites}, permite-nos escrever
\begin{equation*}
\frac{\sqrt{2}+y_3}{4}<\tan(\alpha_2) <
\frac{y_3\left(12+y_3^2\right)}{16} \Rightarrow
\frac{\sqrt{2}+y_3}{4} <
\frac{y_3\left(12+y_3^2\right)}{16}
\text{,}
\end{equation*}
de que resulta a inequação
\begin{equation*}
y_3^3+ 8y_3+ 4\sqrt{2}>0
\text{.}
\end{equation*}
Como o polinómio da inequação tem derivada positiva e admite uma única
raíz real, concluímos de imediato que a mesma constituiu um limite inferior
para $y_3$, sendo esse limite
\begin{equation}
\label{eq:y3min}
y_3^*=\frac{1}{3}\left(54\sqrt{2}+6\sqrt{546}\right)^\frac{1}{3}
-\frac{8}{\left(54\sqrt{2}+6\sqrt{546}\right)^\frac{1}{3}}
\simeq 0.670 \text{.}
\end{equation}
 
Resta-nos determinar $y_1^*$, um minorante para o valor de $y_1$ --- ordenada
onde ocorre a primeira reflexão.
Para o efeito recorremos à ilustração (b) da Figura~\ref{fig:parabOpt3colB},
que nos dá uma representação mais pormenorizada da parte
da cavidade onde ocorrem as duas primeiras reflexões, $B_1$ e $B_2$.
O esquema apresentado foi construído
contando que a primeira reflexão ($B_1$) ocorre num ponto mais elevado
do que o da terceira reflexão ($B_3$). É de facto essa a situação. Isso mesmo pode ser comprovado mostrando
que $\alpha_2$ é sempre menor que o ângulo formado entre o vector normal à curva em $B_2$ 
e o eixo vertical, ou seja
\begin{equation*}
\alpha_2<\frac{\pi}{2}-\arctan\left(\frac{1}{2}y_2\right)
\text{.}
\end{equation*}
Pegando, em \eqref{eq:tanAlpha2Limites}, no limite superior de $\tan(\alpha_2)$ e tendo presente que $y<\sqrt{2}$, construímos a seguinte sequência de desigualdades que comprova o que se pretende:
\begin{equation*}
\alpha_2 < \arctan \left(\frac{y_3\left(12+y_3^2\right)}{16}\right)
< \overbrace{
\arctan \left(\frac{7\sqrt{2}}{8}\right)
}^{\simeq 51.06^\circ}
< \overbrace{\frac{\pi}{2}-\arctan\left(\frac{\sqrt{2}}{2}\right)}^{\simeq 54.74^\circ}
<\frac{\pi}{2}-\arctan\left(\frac{1}{2}y_2\right)
\text{.}
\end{equation*}

Tentemos agora encontrar $y_1^*$.
Podemos definir $y_1^*$ como sendo a ordenada do ponto de intercepção
da parábola esquerda com a semi-recta de origem no ponto $B_2$,
posicionado o mais abaixo
possível ($y_2=y_2^*$),
e com declive igual ao maior valor permitido para o declive
da trajectória que antecede $B_2$ ($\overrightarrow{B_1B_2}$).
A equação da recta que liga $B_1$ a $B_2$ toma a forma
\begin{equation*}
x=m(y-y_2)+x_2\text{,}
\end{equation*}
com $m=\tan(\alpha_1)$ e $x_2=-\frac{1}{4}y_2^2+\frac{1}{2}$. Como estamos interessados
em encontrar o ponto de intersecção dessa recta com a curva parabólica situada no
lado esquerdo, de equação
\begin{equation*}
x=\frac{1}{4}y^2-\frac{1}{2}
\text{ ,}
\end{equation*}
teremos que resolver a equação de segundo grau, na variável $y$, que resulta
da eliminação da variável $x$ por combinação das duas equações anteriores.
Embora tenha duas raízes reais positivas, apenas estamos interessados
na menor delas, que toma a forma
\begin{equation}
\label{eq:y1}
y_1 = 2 m-\sqrt{4 m^2-4 m y_2 - y_2^2+4}
\text{ .}
\end{equation}

Como dissemos, se fizermos $y_2=y_2^*$ e impusermos o declive máximo para a recta,
que nas equações anteriores equivale a considerarmos $m$ mínimo, obtemos $y_1 = y_1^*$.
Sendo $m=\tan(\alpha_1)$, devemos determinar o valor de $\alpha_1$ através do sistema
de equações
\begin{equation*}
\left\{
\begin{array}{l}
\alpha_1 =2\theta_2 + \alpha_2\\
\arctan \left(\frac{1}{2}y_2\right)+\alpha_2 +\theta_2 = \frac{\pi}{2}
\end{array}
\right.
\end{equation*}
que se retira da geometria da
ilustração (b) da Figura~\ref{fig:parabOpt3colB}. Obtem-se
\begin{equation}
\label{eq:alpha1}
\alpha_1 =\pi-2\arctan \left(\frac{y_2}{2}\right)-\alpha_2
\text{.}
\end{equation}
Desta última igualdade, de \eqref{eq:tanAlpha2Limites}, e dado que $y_2<\sqrt{2}$ e $y_3<\sqrt{2}$, deduzimos que
\begin{equation*}
\alpha_1 > \pi-2\arctan \left(\frac{y_2}{2}\right)-\arctan \left(\frac{y_3\left(12+y_3^2\right)}{16}\right)
> \pi-2\arctan \left(\frac{\sqrt{2}}{2}\right)-\arctan \left(\frac{7}{8}\sqrt{2}\right)
\text{,}
\end{equation*}
logo,
\begin{equation*}
m=\tan(\alpha_1) > \tan\left(
\pi-2\arctan \left(\frac{\sqrt{2}}{2}\right)-\arctan \left(\frac{7}{8}\sqrt{2}\right)\right)=
\frac{23}{20}\sqrt{2}
\text{.}
\end{equation*}
Se em \eqref{eq:y1} fizermos $m=\frac{23}{20}\sqrt{2}$ e $y_2 = y_2^*$
obtemos então o limite inferior para $y_1$
\begin{equation}
\label{eq:y1min}
y_1^*=
\frac{23}{10}\sqrt{2}
-\frac{1}{90}\sqrt{
444498-33120\sqrt{2}\sqrt{-51+6\sqrt{79}}
-38400\sqrt{79}
}
\simeq 1.274 \text{.}
\end{equation}
Resumindo,
\begin{equation*}
(y_1^*,y_2^*,y_3^*,y_4^*)\simeq
(1.274, 1.356, 0.670, 0)
\text{.}
\end{equation*}

Com o auxílio da ilustração (b) da Figura~\ref{fig:parabOpt3colB} vamos,
por fim, analisar o ângulo de entrada $\varphi$ da partícula. Com o sistema
de equações ($\frac{1}{2}y_1=\frac{\dd x}{\dd y}\left.\right|_{y=y_1}$)
\begin{equation*}
\left\{
\begin{array}{l}
2\theta_1+\varphi +\alpha_1 = \pi\\
\arctan \left(\frac{1}{2}y_1\right)+\varphi +\theta_1 = \frac{\pi}{2}
\end{array}
\right.
\end{equation*}
e com as igualdades \eqref{eq:alpha1} e \eqref{eq:alpha2b}
obtemos, sucessivamente,
\begin{eqnarray}
\varphi \hspace{-0.25cm}&=&\hspace{-0.25cm}\alpha_1-2\arctan \left(\frac{y_1}{2}\right)
\text{,}\nonumber\\
\label{eq:phiP2}
\varphi \hspace{-0.25cm}
&=&\hspace{-0.25cm}
\pi-2\arctan \left(\frac{y_1}{2}\right)
-2\arctan \left(\frac{y_2}{2}\right)-\alpha_2
\text{,}\\
\varphi \hspace{-0.25cm}
&=&\hspace{-0.25cm}
\pi-2\arctan \left(\frac{y_1}{2}\right)
-2\arctan \left(\frac{y_2}{2}\right)
-2\arctan \left(\frac{y_3}{2}\right)+\arctan \left(\frac{y_3+y_4}{4}\right)
\text{.}\hspace{1.0cm}
\nonumber
\end{eqnarray}

Pegando em \eqref{eq:phiP2}, de \eqref{eq:alpha2Maj} deduzimos que
\begin{equation*}
\varphi < \pi-2\arctan \left(\frac{y_1}{2}\right)
-2\arctan \left(\frac{y_2}{2}\right)
-\arctan \left(\frac{\sqrt{2}+y_3}{4}\right)
\text{.}
\end{equation*}
De acordo com as definições \eqref{eq:minorantes} e com os valores encontrados em \eqref{eq:y2min}, \eqref{eq:y3min} e \eqref{eq:y1min}, podemos concluir que
\begin{equation*}
\varphi < \pi-2\arctan \left(\frac{y_1^*}{2}\right)
-2\arctan \left(\frac{y_2^*}{2}\right)
-\arctan \left(\frac{\sqrt{2}+y_3^*}{4}\right)
\simeq 19.18^\circ
\text{,}
\end{equation*}
ou seja
\begin{equation*}
\varphi < \varphi_0 \simeq 19.47^\circ
\text{.}
\end{equation*}
Com isto podemos finalmente concluir que é impossível termos uma quarta reflexão,
pois, a acontecer, a partícula teria que ter entrado na cavidade com um
ângulo $\varphi < \varphi_0$, como acabámos de mostrar --- algo
que contraria a nossa imposição inicial, $\varphi > \varphi_0$.
Como a cavidade apresenta simetria em relação ao seu eixo vertical central,
a conclusão a que chegamos é igualmente válida para $\varphi < -\varphi_0$,
ficando assim provado o que pretendíamos (Teorema~\ref{teor:3ref}):
\begin{quotation}{\sl
Para $\left|\varphi\right|>\varphi_0$, ocorrem sempre três reflexões, alternadamente nas faces esquerda e direita da cavidade Dupla Parábola.
}\end{quotation}


\subsection{Número mínimo de reflexões}
\label{cha:min3col}

\begin{theorem}
\label{teor:min3ref}
Qualquer partícula que entre na cavidade Dupla Parábola descreve uma trajectória com um mínimo de 3 reflexões.
\end{theorem}

Para $\left|\varphi\right|>\varphi_0$ o postulado está já demonstrado, porque sabemos, da prova anterior, que nesse caso as trajectórias são sempre de 3 reflexões.
Faltará então demonstrar para $-\varphi_0<\varphi<\varphi_0$, mas dada a simetria da cavidade apenas teremos necessidade de provar a veracidade da afirmação para o intervalo $0<\varphi<\varphi_0$.

Basearemos a nossa demonstração em algumas das deduções que fizemos na prova anterior (\S\ref{cha:condSuf3col}), sendo-nos especialmente úteis as ilustrações da Figura~\ref{fig:parabOpt3col}.
Assumiremos ainda como verdadeira a seguinte premissa: 
``Se a 2ª reflexão acontecer na mesma face da cavidade onde ocorreu
a 1ª reflexão, haverá necessariamente uma terceira reflexão''.
Dispensamo-nos de provar este princípio por nos parecer evidente.

Para $0<\varphi<\varphi_0$ a primeira reflexão tanto pode ocorrer na face do lado esquerdo
como na do lado direito. Analisemos cada um dos casos em separado.

\vspace{0.5cm}
\noindent
\emph{1ª reflexão no lado esquerdo}

Sendo $0<\varphi<\varphi_0$ poderemos ter as duas primeiras reflexões na face do lado esquerdo,
estando neste caso garantida, como assumimos atrás, a existência de 3 ou mais reflexões.
Caso contrário, sendo a segunda reflexão no lado direito, uma parte inicial da trajectória poderá ser sempre representada pelas primeiras três ilustrações da Figura~\ref{fig:parabOpt3col} (assumindo $0<\varphi<\varphi_0$), as quais garantem, também nesse caso, a existência de uma terceira reflexão $B_3$.
Para provarmos o que acabámos de dizer bastar-nos-á demonstrar a natureza ascendente do subtrajecto $\overrightarrow{B_1B_2}$.

Fixemos na parábola do lado esquerdo (ilustração~(a) da Figura~\ref{fig:parabOpt3col}) o primeiro ponto de reflexão $B_1$. Para qualquer que seja $B_1$ é sempre possível traçarmos um subtrajecto inicial $\overrightarrow{B_0B_1}$ com origem num ângulo de entrada $\varphi>\varphi_0$. Como em \S\ref{cha:condSuf3col} (pág.~\pageref{pg:subtrajB1B2})
mostrámos ser ascendente o subtrajecto $\overrightarrow{B_1B_2}$ que se seguiria, o mesmo necessariamente sucederá para qualquer que seja $0<\varphi<\varphi_0$, dado que neste caso $\overrightarrow{B_0B_1}$ apresentará um declive negativo mais acentuado.
Uma vez que em \S\ref{cha:condSuf3col} (pág.~\pageref{pg:subtrajB2B3})
caracterizámos
$\overrightarrow{B_2B_3}$ apenas com base na natureza ascendente do subtrajecto precedente $\overrightarrow{B_1B_2}$, as conclusões a que chegámos para
$\overrightarrow{B_2B_3}$ são igualmente válidas para $0<\varphi<\varphi_0$.


\vspace{0.5cm}
\noindent
\emph{1ª reflexão no lado direito}

\begin{figure}[!hb]
\begin{center}
\includegraphics*[width=0.34\columnwidth]{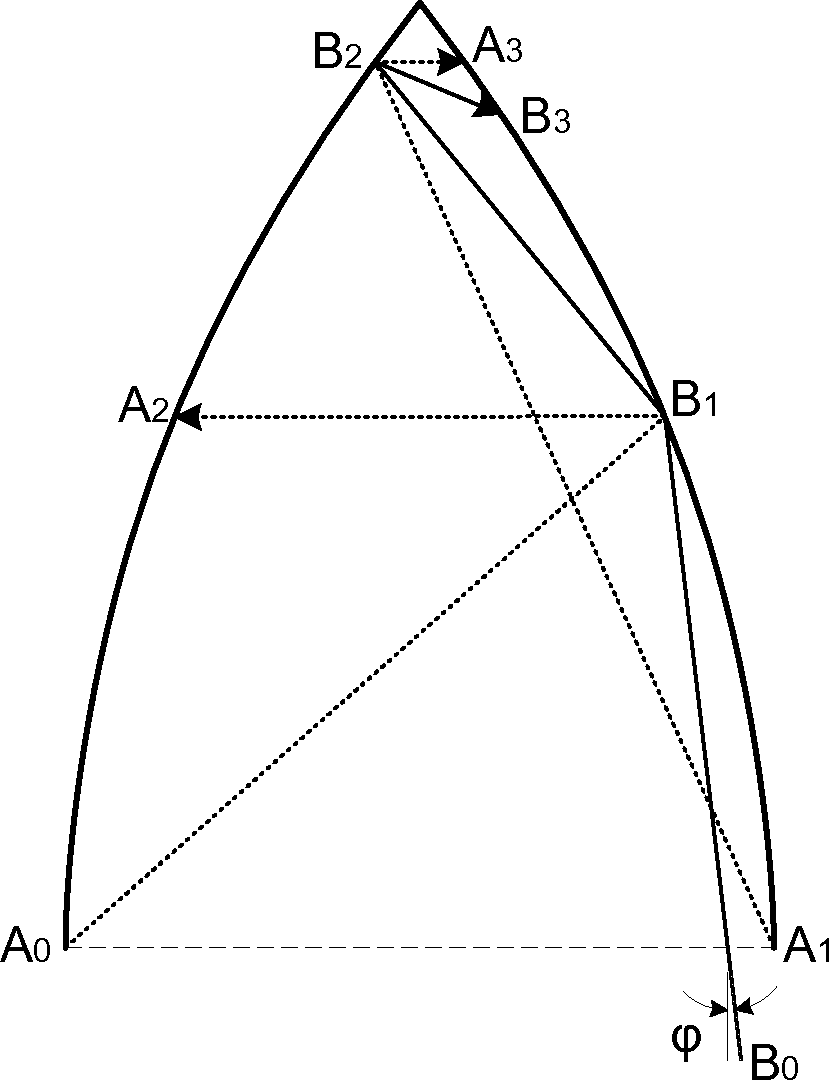}
\caption[Ilustração para estudo das trajectórias com
ângulos de entrada $0<\varphi<\varphi_0$.]{Esquema ilustrativo para estudo da trajectória de partículas com ângulos de entrada $0<\varphi<\varphi_0$.}
\label{fig:parabOptMin3col}
\end{center}
\end{figure}

Também neste caso poderemos ter as duas primeiras reflexões na face do lado direito,
ficando garantida a existência de 3 ou mais reflexões.
Se isso não acontecer, teremos necessariamente uma trajectória com o
aspecto da trajectória $B_0B_1B_2B_3$ ilustrada no esquema
da Figura~\ref{fig:parabOptMin3col},
onde surgem igualmente representados dois trajectos auxiliares (a tracejado),
$A_0B_1A_2$ e $A_1B_2A_3$, que,
ao passarem pelos focos das parábolas, apresentam o
subtrajecto posterior à reflexão horizontal.
Tendo por base as leis de reflexão, podemos sucintamente deduzir o seguinte:
como o ângulo $\widehat{A_0B_1A_2}$ tem que ser interior ao ângulo
$\widehat{B_0B_1B_2}$, concluímos que $\overrightarrow{B_1B_2}$ é 
de natureza ascendente; 
como $\widehat{B_1B_2B_3}$ é necessariamente um ângulo interior a $\widehat{A_1B_2A_3}$,
concluímos que $B_3$ tem que se situar entre $A_1$ e $A_3$, o que garante
a existência da terceira reflexão.
Fica assim concluída a demonstração do Teorema~\ref{teor:min3ref}.

\section{Outras possíveis aplicações}
\label{sec:outrasAplic}

Para além de maximizar a resistência newtoniana,
é entusiasmante verificar que as potencialidades
da forma Dupla Parábola por nós encontrada, podem
vir a revelar-se também muito interessantes noutros domínios 
de interesse prático.

Se revestirmos a parte interior da cavidade Dupla Parábola com uma ``superfície'' polida,
a trajectória da luz no seu interior será descrita pelos princípios da óptica geométrica, designadamente propagação rectilínea da luz, leis de reflexão
e reversibilidade da luz. Assim, os modelos computacionais que foram por nós desenvolvidos para simular a dinâmica de bilhar no interior da cada uma das formas estudas (onde se consideraram colisões de partículas perfeitamente elásticas) são igualmente válidos quando o problema passa a ser de natureza óptica. Olhemos estão para a forma 2D por nós encontrada nessa
nova perspectiva.

Dadas as características de reflexão que apresenta a forma Dupla Parábola,
rapidamente vislumbramos-lhe uma propensão natural
para poder vir a ser usada com grande sucesso
no desenho de retrorreflectores.
Retrorreflectores são dispositivos que enviam a luz ou outra
radiação incidente de volta à fonte emissora. Idealmente,
o retrorreflector deve desempenhar essa função
independentemente do ângulo de incidência, algo que
não acontece com os dispositivos existentes.
Nesses dispositivos a superfície retrorreflectora é composta
por uma unidade óptica de reflexão ou, na maior parte dos casos,
por um conjunto considerável desses elementos em tamanho reduzido.
Esses elementos ópticos individuais --- responsáveis pela
inversão do fluxo incidente --- podem ser
pequenos elementos esféricos
com propriedades ópticas adequadas, mas o mais habitual é serem pequenos
cantos cúbicos retrorreflectores --- cantos formados por três espelhos
mutuamente perpendiculares ---,
que como se sabe apenas invertem a radiação para determinados
ângulos de incidência.

Como vimos, a Dupla Parábola, embora não garanta a inversão perfeita de toda a
radiação incidente, desempenha essa função com
grande sucesso (os pontos $(\varphi,\varphi^+)$ concentram-se nas proximidades da diagonal $\varphi=\varphi^+$, Figura~\ref{fig:Dist_Phi_PhiPlus}):
garante uma grande aproximação das direcções dos fluxos
incidente e reflectido para uma parte significativa dos ângulos de incidência,
e mesmo para os restantes não permite que o desfazamento atinja valores
elevados. Antevemos, por isso, bastante promissora a
sua possível utilização na definição de novas geometrias para
os elementos ópticos que compõem as superfícies retrorreflectoras.

Os dispositivos retrorreflectores, embora sejam usados nas mais diversificadas
áreas tecnológicas, como é o caso por exemplo das comunicações ópticas em espaço
aberto, é na indústria automóvel e sinalização rodoviária que,
sendo utilizados de forma massiva, nos apercebemos mais facilmente da sua utilidade.
Vejamos, na secção que se segue,
uma forma possível de tirarmos partido do nosso resultado.

\subsubsection*{Retrorreflectores na indústria automóvel e sinalização rodoviária}
Hoje em dia todas as principais vias de circulação automóvel dispõem de
dispositivos retrorreflectores, complementares à sinalização horizontal,
instalados fora da superfície pavimentada e a uma altura predeterminada --- designados delineadores ---, com o objectivo de sinalizar de forma clara a geometria da via, mesmo perante condições de visibilidade reduzida como é o caso da condução nocturna ou sob condições climatéricas adversas.
\begin{figure}[!hb]
\begin{center}
\includegraphics*{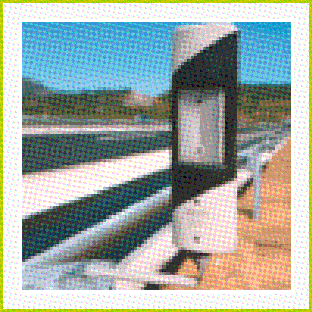}
\caption[Exemplo de retrorreflector do tipo delineador.]{Exemplo de retrorreflector do tipo delineador (imagem retirada do sítio Web da BRISA --- Auto-estradas de Portugal).}
\label{fig:delineador}
\end{center}
\end{figure}
Cientes do importante papel que esse tipo de sinalização desempenha
no aumento dos níveis de segurança rodoviária, tentamos, com base nos resultados obtidos, 
propor um esquema diferente de dispositivos retrorreflectores que, reflectindo
a luz de forma mais eficiente, apresentarão certamente coeficientes de intensidade luminosa mais elevados (indicador normalmente usado para medir o desempenho dos retrorreflectores).

\begin{figure}[!ht]
\begin{center}
\includegraphics*[width=0.5\columnwidth]{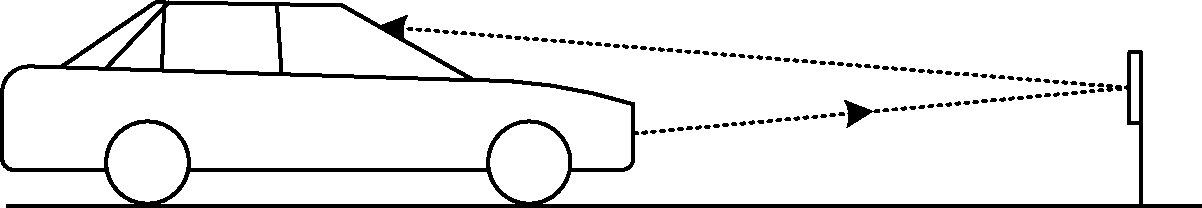}
\caption[Esquema da luz reflectida por dispositivos com
superfície reflectora vertical.]{Esquema ilustrativo da luz reflectida por dispositivos com
superfície reflectora vertical.}
\label{fig:reflectorEstrada6}
\end{center}
\end{figure}

\begin{figure}[!hb]
\begin{center}
\includegraphics*[width=0.5\columnwidth]{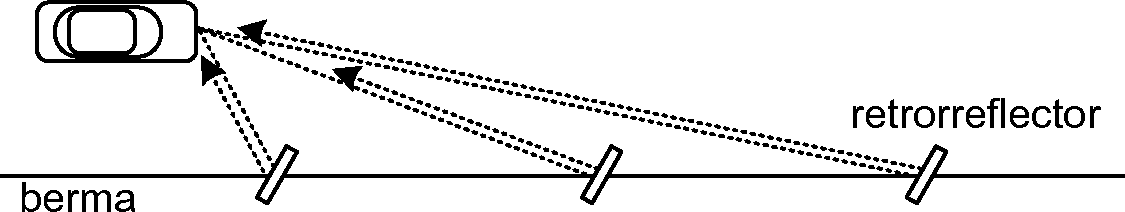}
\caption{Esquema ilustrativo da luz reflectida por retrorreflectores ideais.}
\label{fig:reflectorEstrada5}
\end{center}
\end{figure}

\begin{figure}[!hb]
\begin{center}
\includegraphics*[width=0.45\columnwidth]{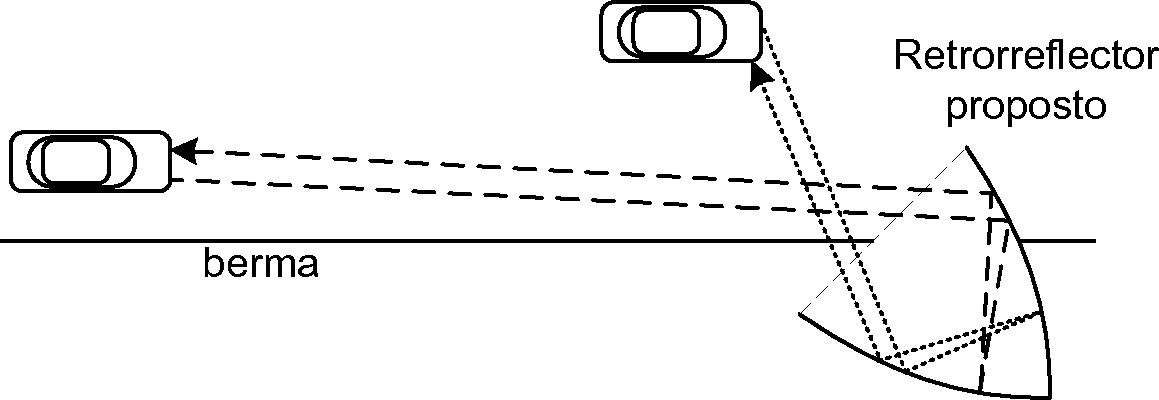}
\caption{Esquema ilustrativo da luz reflectida pelo retrorreflector proposto.}
\label{fig:reflectorEstrada8}
\end{center}
\end{figure}

\begin{figure}[!ht]
\begin{center}
\includegraphics*[width=0.25\columnwidth]{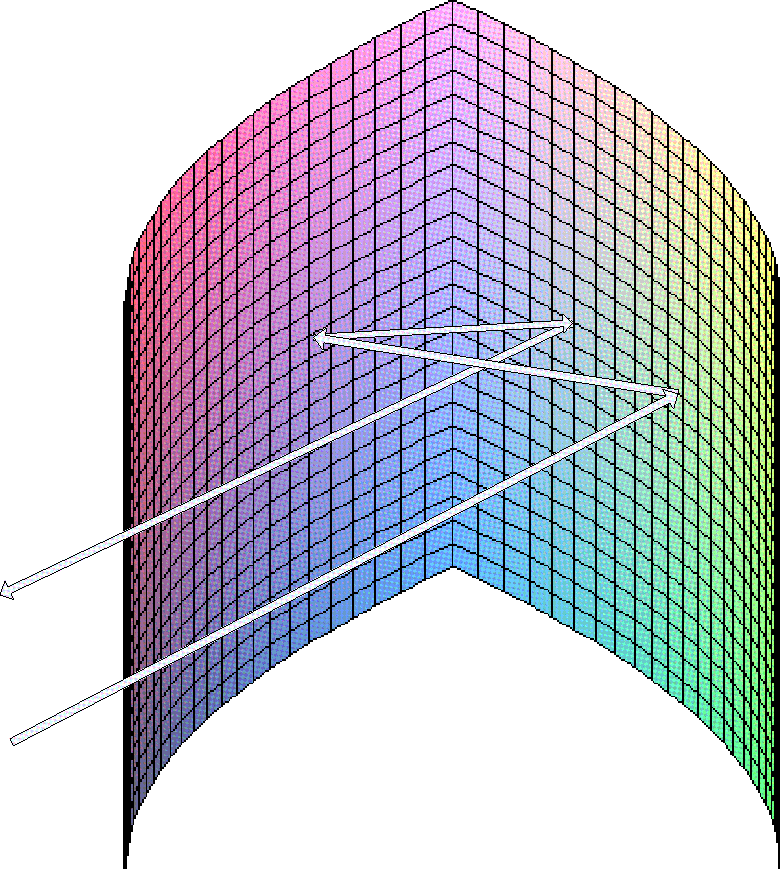}
\caption{Forma 3D da superfície proposta para elemento retrorreflector.}
\label{fig:reflectorEstrada4}
\end{center}
\end{figure}

Embora a forma por nós encontrada tenha existência no espaço bidimensional,
é possível, para determinadas aplicações práticas, construir
a partir dela formas 3D que mantenham as propriedades ópticas adequadas.
Vejamos o caso dos retrorreflectores usados como delineadores nas vias de comunicação.
Como a altura a que se encontram do pavimento, tanto 
os olhos do condutor (ponto de observação) como os faróis do veículo (fonte de luz),
é sensivelmente a mesma para a generalidade dos automóveis, pode-se facilmente
projectar um reflector que garanta que o raio de luz reflectido,
quando projectado no plano vertical que une o veículo ao reflector,
passe ao nível do ponto de observação.
Esse efeito é conseguido com espelhos de superfície
vertical colocados a uma altura equivalente
à média das alturas a que se encontram o ponto
de observação e a fonte de luz --- aproximadamente $75$~cm acima do pavimento da faixa de rodagem ---,
tal como ilustrado na figura~\ref{fig:reflectorEstrada6}.
Assim, a parte da geometria da superfície reflectora
que falta ainda definir é neste caso a forma da sua
projecção no plano horizontal. Será essa forma que determinará
o grau de aproximação dos raios de luz incidente e reflectido, quando
projectados num plano horizontal, e que idealmente deverão ter
a mesma direcção --- ver figura~\ref{fig:reflectorEstrada5}.
Tratando-se de uma curva 2D,
a forma Dupla Parábola será seguramente a melhor escolha,
pois é aquela que em nosso entender apresenta melhor desempenho --- ver figura~\ref{fig:reflectorEstrada8}.
Assim, a superfície retrorreflectora 3D que propomos é a parte interior
da superfície varrida pelo deslocamento vertical do contorno da nossa forma 2D
contida num plano horizontal,
tal como se ilustra na figura~\ref{fig:reflectorEstrada4}.
Um inovador dispositivo retrorreflector pode então ser construído integrando
o elemento proposto como unidade única de reflexão ou um conjunto desses
elementos distribuídos numa área rectangular segundo uma disposição matricial.

Se se entender que também no plano vertical o dispositivo deve
reflectir a luz de forma a ter a mesma direcção da luz incidente,
a solução passará por se introduzir uma pequena alteração na forma
do elemento retrorreflector: para além da superfície vertical,
a forma passará a ser delimitada no topo e na sua base
por duas superfícies planas horizontais.
Sabemos que, com esta configuração, o sentido da luz é invertido sempre que 
ocorra um número ímpar de reflexões nas superfícies horizontais ---
lados superior e inferior. Isto garante-nos que uma boa parte da luz reflectida
toma efectivamente a direcção da luz incidente.
Em \cite{gouvPhD} foram já realizadas algumas simulações numéricas para o caso 3D,
tendo os resultados apontado para o bom desempenho desta configuração --- 
trata-se da forma 3D de maior resistência que se conseguiu obter.

\begin{figure}[!ht]
\begin{center}
\includegraphics*[width=0.45\columnwidth]{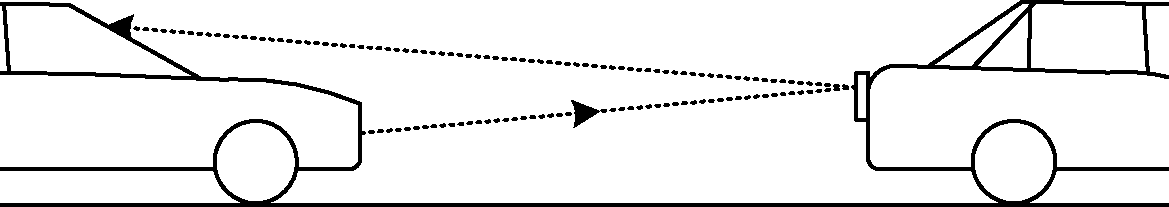}
\caption{Esquema da reflexão de luz entre dois veículos.}
\label{fig:reflectorEstrada7}
\end{center}
\end{figure}

O mesmo tipo de solução pode ser usado para os
retrorreflectores instalados na parte posterior das viaturas.
Também neste caso o reflector encontra-se normalmente a uma altura intermédia
entre a fonte de luz e o ponto de observação, como se ilustra na figura~\ref{fig:reflectorEstrada7}.

Poderão também ser conseguidos óptimos resultados
com o uso combinado do nosso elemento reflector com outros
reflectores já existentes, dado que se podem complementar.
Por exemplo, as formas em ângulo recto têm um óptimo desempenho para
ângulos de incidência baixos, mas deixam de ter qualquer eficácia para ângulos
superiores a $45^\circ$; pelo contrário, o nosso elemento reflector tem um desempenho
irrepreensível para ângulos superiores a $45^\circ$,
e perde alguma eficácia em ângulos de incidência de baixa amplitude.

Ainda que as reais mais-valias do nosso modelo de retrorreflexão apenas possam ser
verdadeiramente comprovadas através da realização de ensaios em laboratórios especializados
que permitam a análise comparativa com outros modelos,
estamos em todo o caso seguros da sua eficácia como retrorreflector, em situações
em que os emissores/receptores se movimentem num mesmo plano, ou então,
se quisermos ser mais precisos, em todas as situações em que seja possível
encontrar um plano bidimensional que intercepte perpendicularmente
e a meio o eixo que une o emissor ao receptor de todos os
pares emissor/receptor admissíveis.
Nos exemplos que demos considerámos ser essa a situação, embora saibamos
que na prática não se passe exactamente assim.


\section{Conclusão}
\label{sec:concl}
\begin{figure}[!ht]
\begin{center}
\includegraphics[width=0.3\columnwidth]{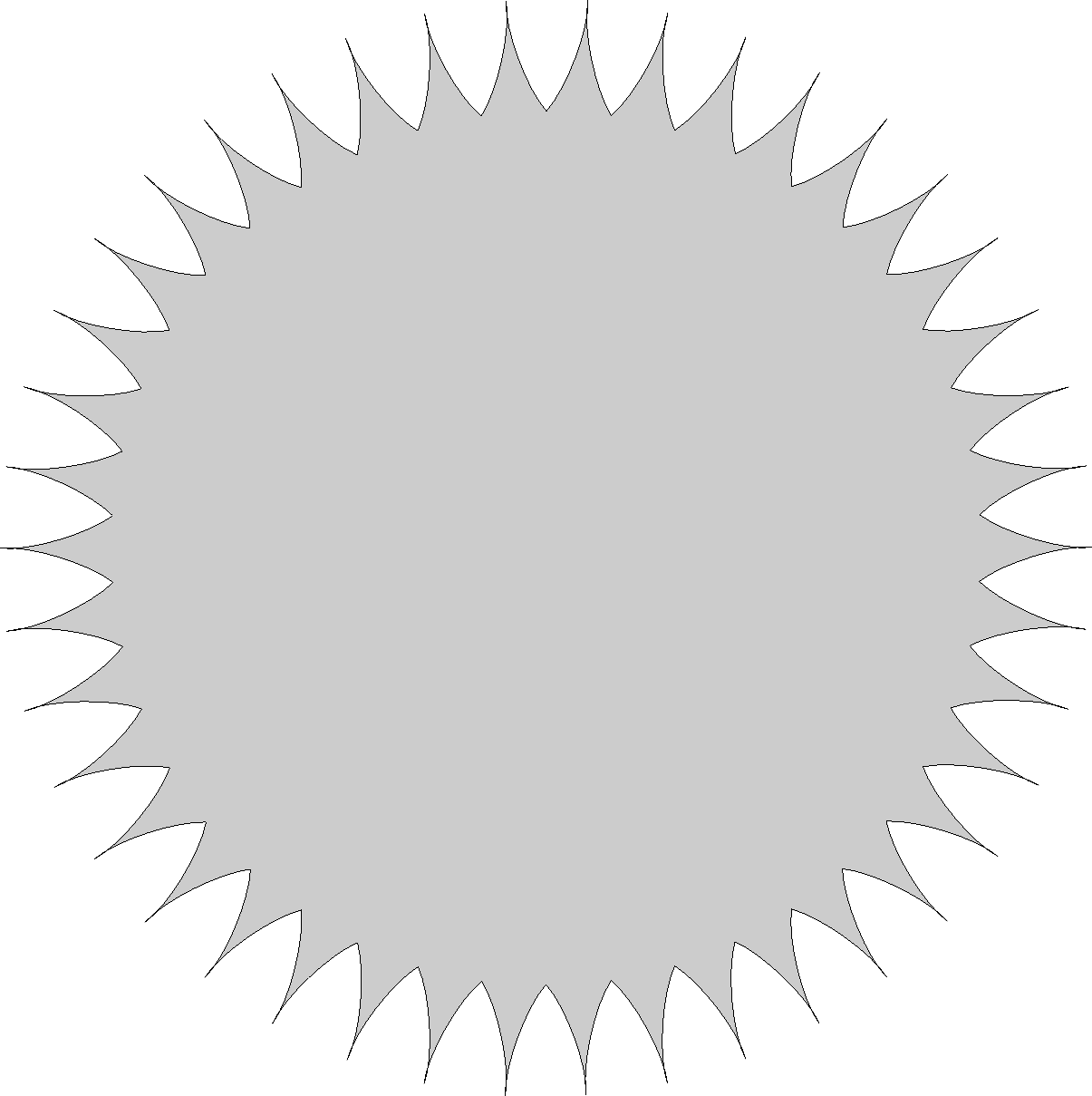}
\caption{Corpo 2D (quase) óptimo.}
\label{fig:discoOptimo}
\end{center}
\end{figure}

Na continuação do estudo realizado anteriormente pelos autores em \cite{Plakhov07:CM, Plakhov07}, conseguiu-se, com o trabalho agora apresentado, obter 
um resultado original que nos parece de grande alcance: os algoritmos de optimização convergiram para uma
forma geométrica muito próxima da forma ideal --- a \emph{Dupla Parábola}. Trata-se de uma forma de rugosidade que confere uma resistência quase máxima (muito próxima do majorante teórico) a um disco que, para além de se deslocar num movimento translacional, rode lentamente sobre si mesmo.
Na Figura~\ref{fig:discoOptimo} é mostrado um desses corpos.
Atendendo a que o contorno do corpo apresentado é integralmente formado por $42$ cavidades $\Omega$ com a forma da Dupla Parábola, cada uma das quais com uma resistência relativa
de $1.49650$, de \eqref{eq:RB2} e \eqref{eq:RxPerimetros} concluímos que R(B)=$\frac{\sin(\pi/42)}{\pi/42}R(\Omega)\approx 1.4951$ é a resistência
total do corpo, um valor
$49.51\%$ acima do valor de resistência do disco de contorno liso correspondente
(o menor disco que inclua o corpo).
Sabemos que se o corpo for formado por um número suficientemente elevado de pequenas dessas cavidades, a sua resistência atingirá mesmo o valor $1.4965$,
mas o exemplo apresentado é suficiente para percebermos o quão próximos ficámos
do conhecido majorante teórico ($50\%$).

Ainda que o valor da resistência da Dupla Parábola tenha sido determinado numericamente,
o estudo que se realizou em \S\ref{sec:caract}, com o objectivo de se caracterizar analiticamente as reflexões no interior dessa forma de cavidade, consolida os resultados apresentados. 
Conseguimos demonstrar algumas propriedades importantes que ajudam a perceber o elevado valor de resistência que se obteve. Tentaremos no futuro desenvolver outros estudos teóricos que venham a permitir consolidar ainda mais este resultado.
Por exemplo, um problema interessante em aberto consiste em delimitar o desfazamento entre
os ângulos de entrada e de saída para as trajectórias de 3 reflexões ---
para as restantes (trajectórias com 4 ou mais reflexões) sabemos já que $\left|\varphi-\varphi^+\right|<2\varphi_0 \simeq 2 \times 19.47^\circ$.

A Dupla Parábola é efectivamente um resultado de grande alcance prático.
Para além de permitir elevar a resistência newtoniana quase ao seu majorante teórico,
essa forma geométrica revela-se também muito interessante noutros domínios 
de aplicação.
Dadas as suas características de reflexão,
rapidamente vislumbramos-lhe uma propensão natural
para poder vir a ser usada com grande sucesso
no desenho de retrorreflectores --- veja-se em \S\ref{sec:outrasAplic} o estudo exploratório sobre a sua possível utilização
na sinalização rodoviária e indústria automóvel. 
Mas a sua utilidade estende-se a muitos outros domínios de aplicação importantes, como será certamente o caso dos sistemas de comunicações ópticas em espaço aberto ou dos sistemas ópticos que integram alguns dos equipamentos tecnológicos ligados às ciências da saúde e à indústria em geral.
Na continuação deste trabalho, tencionamos vir também a desenvolver investigações
que permitam a análise comparativa de superfícies baseadas na Dupla Parábola com outros modelos de retrorreflexão existentes. Nesse estudo usaremos, em substituição da função resistência, uma função de custo mais adequada à avaliação das capacidades retrorreflectivas --- uma função que contabilize devidamente o desfazamento angular entre os fluxos incidente e reflectido.

Uma incursão no domínio tridimensional, realizada em \cite{gouvPhD}, mostrou também ela, ser a Dupla Parábola uma forma de cavidade muito especial. A nossa convicção no seu virtuosismo saiu realmente reforçada quando obtivemos o melhor resultado para o caso 3D. Esse resultado foi conseguido com uma cavidade cuja superfície é a área varrida pelo deslocamento da curva Dupla Parábola na direcção perpendicular ao seu plano. 
Tendo o valor da sua resistência ($R=1.80$) ficado um pouco aquém do majorante teórico para o caso 3D ($R=2$), superar esse valor será também um desafio interessante a considerar futuramente.

Já para o caso 2D pressentimos maior dificuldade em virmos a superar o resultado já alcançado --- quer pela proximidade a que este se encontra do majorante teórico, quer pelo facto de termos já realizado, sem sucesso, uma série de investigações visando esse objectivo.
Tendo ainda em conta que o majorante $1.5$ apenas significa a não existência de formas que superem esse valor de resistência, a forma por nós encontrada poderá mesmo tratar-se de uma
solução óptima. A confirmar-se esta hipótese --- o valor da resistência que caracteriza a Dupla Parábola tratar-se efectivamente do limite máximo que é possível alcançar com formas bidimensionais reais --- o nosso resultado ganhará ainda uma importância redobrada. 
Embora não se perspectivem fáceis desenvolvimentos, este é um importante problema que fica em aberto à espera de futuras contribuições que, se não vierem a superar o nosso resultado,
venham a permitir reforçar a nossa conjectura.


\section*{Agradecimentos}
A investigação dos autores teve o apoio do \textsf{Centro de Estudos em Optimização e Controlo} (CEOC), Unidade de Investigação e Desenvolvimento (UI\&D) da Universidade de Aveiro, financiada pela FCT. A investigação do primeiro autor foi ainda co-financiada pelo \textsf{Fundo Social Europeu}, através do Programa \textsf{PRODEP III/5.3/2003}.

\bibliographystyle{plain}
\bibliography{../../../tesegouv/resist}



\end{document}